\documentstyle[psfig,aps]{revtex}

\begin{document}
\title{Evidence for structural and electronic instabilities at
intermediate temperatures in $\kappa$-(BEDT-TTF)$_{2}$X for
X=Cu[N(CN)$_{2}$]Cl, Cu[N(CN)$_{2}$]Br and Cu(NCS)$_{2}$:
Implications for the phase diagram of these quasi-2D organic
superconductors}
\author{J.\ M\"uller}
\address{Max-Planck-Institut f\"ur Chemische Physik fester Stoffe, D-01187
Dresden, Germany}
\author{M.\ Lang}
\address{Physikalisches Institut der Universit\"at Frankfurt, D-60054
Frankfurt am Main, Germany}
\author{F.\ Steglich}
\address{Max-Planck-Institut f\"ur Chemische Physik fester Stoffe, D-01187
Dresden, Germany}
\author{J.\ A.\ Schlueter and A.\ M.\ Kini}
\address{Materials Science Division, Argonne National Laboratory, Argonne, Illinois, USA}
\author{T.\ Sasaki}
\address{Institute for Materials Research, Tohoku University, Sendai, Japan}

\date{\today}
\maketitle

\begin{abstract}
We present high-resolution measurements of the coefficient of
thermal expansion $\alpha (T)=\partial \ln l(T)/\partial T$ of the
quasi-twodimensional (quasi-2D) salts $\kappa$-(BEDT-TTF)$_2$X
with X = Cu(NCS)$_2$, Cu[N(CN)$_2$]Br and Cu[N(CN)$_2$]Cl in the
temperature range $T \leq 150$\,K. Three distinct kinds of
anomalies corresponding to different temperature ranges have been
identified. These are (A) phase-transition anomalies into the
superconducting, X = Cu(NCS)$_2$, Cu[N(CN)$_2$]Br, and
antiferromagnetic, X = Cu[N(CN)$_2$]Cl, ground state, (B)
phase-transition-like anomalies at intermediate temperatures (30 -
50)\,K for the superconducting salts and (C) kinetic, glass-like
transitions at higher temperatures, i.e.\ (70 - 80)\,K for all
compounds. By a thermodynamic analysis of the discontinuities at
the second-order phase transitions that characterize the ground
state of the system (A), the uniaxial-pressure coefficients of the
respective transition temperatures could be determined.  We find
that in contrast to what has been frequently assumed, the
intraplane-pressure coefficients of $T_c$ for this family of
quasi-2D superconductors do not reveal a simple form of
systematics. This demonstrates that attempts to model these
systems by solely considering in-plane electronic parameters are
not appropriate. At intermediate temperatures (B), distinct
anomalies reminiscent of second-order phase transitions have been
found at $T^\ast = 38$\,K and 45\,K for the superconducting X =
Cu(NCS)$_2$ and Cu[N(CN)$_2$]Br salts, respectively. Most
interestingly, we find that the signs of the uniaxial pressure
coefficients of $T^\ast$, $\partial T^\ast /
\partial p_i$ $(i = a,b,c)$, are strictly anticorrelated with those
of $T_c$. Based on comparative studies including the
non-superconducting X = Cu[N(CN)$_2$]Cl salt as well as
isotopically labeled compounds, we propose that $T^\ast$ marks the
transition to a spin-density-wave (SDW) state forming on minor,
quasi-1D parts of the Fermi surface. Our results are compatible
with two competing order parameters that form on disjunct portions
of the Fermi surface. At elevated temperatures (C), all compounds
show $\alpha (T)$ anomalies that can be identified with a kinetic,
glass-like transition where, below a characteristic temperature
$T_g$, disorder in the orientational degrees of freedom of the
terminal ethylene groups becomes frozen in. We argue that the
degree of disorder increases on going from the X = Cu(NCS)$_2$ to
Cu[N(CN)$_2$]Br and the Cu[N(CN)$_2$]Cl salt. Our results provide
a natural explanation for the unusual time- and cooling-rate
dependencies of the ground-state properties in the hydrogenated
and deuterated Cu[N(CN)$_2$]Br salts reported in the literature.
\end{abstract}

\pacs{PACS numbers: 74.70.Kn}%, 65.70.+y, 74.25.Bt}

\twocolumn{\hsize\textwidth\columnwidth\hsize\csname\endcsname
\draft}

\narrowtext

\section{Introduction}

Organic charge-transfer salts based on the electron-donor molecule
bis(ethylenedithio)-tetra\-thia\-fulvalene, commonly abbreviated
BEDT-TTF or simply ET, are characterized by their
quasi-twodimensional electronic properties. They consist of
alternating conducting ET layers and insulating anion sheets.
Within the conducting layers, the delocalization of the charge
carriers is provided by the overlap of the $\pi $ orbitals of
sulfur atoms of adjacent ET molecules. The packing pattern of the
ET molecules, and, thereby the electronic properties, are
determined by the anion structure via short C-H\thinspace $\cdots
$\thinspace anion contacts of the terminal ethylene groups of the
ETs. In forming the solid, these [(CH$_{2}$)$_{2}$] endgroups can
adopt two possible out-of-plane configurations, with the relative
orientation of the outer C-C bonds being either {\em eclipsed} or
{\em staggered}. While at high temperatures the [(CH$_{2}$)$_{2}$]
system is thermally disordered, one of the two configurations,
depending on the anion and crystal structure, is adopted upon
lowering the temperature. This points to the fact that the
ethylene conformation is an important parameter determining the
structural and physical properties of the ET compounds
\cite{Leung85,Geiser91,Ishiguro97}.

Within this class of materials, the $\kappa $-phase (ET)$_{2}$X
salts are of particular interest because of their interesting
superconducting and normal-state properties - some of which are
similar to those of the high-$T_c$ cuprates
\cite{Lang96,McKenzie97}. The compounds with the complex anions
X$^{-}$=[Cu(NCS)$_{2}$]$^{-}$ and [Cu\{N(CN)$_{2}$\}Br]$^{-}$,
which will be abbreviated as $\kappa $-Cu(NCS)$_{2}$ and $\kappa
$-Br, are superconductors with $T_{c}$ values of 10.4\thinspace K
and 11.6\thinspace K, respectively. On the other hand, the system
$\kappa $-(ET)$_{2}$Cu[N(CN)$_{2}$]Cl, in short $\kappa $-Cl, is
an antiferromagnetic insulator ($T_{N}=27\,{\rm K}$) which can be
transformed into a superconductor with $T_{c}=12.8\,{\rm K}$ upon
applying a small hydrostatic pressure of only 300\thinspace bar
\cite{Wang91}. Kanoda \cite{Kanoda97} proposed a conceptual phase
diagram for the dimeric $\kappa $-type BEDT-TTF salts where the
application of hydrostatic pressure has been linked to a variation
of in-plane electronic parameters only, i.e.\ $U_{{\rm eff}}/W$
with an effective on-site (dimer) Coulomb interaction $U_{{\rm
eff}}$ and a bandwidth $W$: with increasing $U_{{\rm eff}}/W$ the
ground state changes from a paramagnetic metal (PM) to a
superconductor (SC) and further to an antiferromagnetic insulator
(AFI). The positions of the various salts with different anions in
the phase diagram is determined by their ambient-pressure
ground-state
properties. In this picture, the deuterated salt $\kappa $-(D$_{8}$-BEDT-TTF)%
$_{2}$Cu[N(CN)$_{2}$]Br, denoted as $\kappa $-D$_{8}$-Br, is situated right
at the AFI/SC border in between the antiferromagnetic insulating $\kappa $%
-Cl and the superconducting hydrogenated $\kappa $-H$_{8}$-Br
salts. The close proximity of an antiferromagnetic insulating to a
superconducting phase has been considered - in analogy to the
high-$T_c$ cuprates - as a strong indication that both phenomena
are closely connected to each other \cite{McKenzie97}. A
theoretical approach to the above proposal has been given by Kino
and Fukuyama \cite{Kino96} on the basis of a two-dimensional (2D)
Hubbard model. In this picture, the AFI state of $\kappa
$-(ET)$_{2}$X is a Mott insulating phase. The Mott-Hubbard
scenario for the title-compound family implies a half-filled
conduction band together with strong electron correlations leading
to unusual normal-state properties in the metallic phase
(pseudogap behavior) close to the magnetic insulating phase and a
spin-fluctuation mediated superconductivity
\cite{Kanoda97,Nakazawa00}.\newline In fact, various properties of
the normal state show anomalous behavior:
the spin-lattice-relaxation rate $(T_{1}T)^{-1}$ of the superconducting $%
\kappa $-H$_{8}$-Br and $\kappa $-Cu(NCS)$_{2}$ salts studied by $^{13}$%
C-NMR shows a maximum around 50\thinspace K which has been
ascribed to antiferromagnetic spin fluctuations \cite{Kawamoto95}.
For both compounds, a decrease of the spin susceptibility $\chi
_{{\rm Spin}}$ in the same temperature range
\cite{Mayaffre94,Kawamoto95a}, together with a distinct peak in
the temperature derivative of the electrical resistivity, $dR/dT,$
\cite{Sushko91,Gärtner88,Murata90} has been interpreted as a
reduction of the density of states at the Fermi level, i.e.\ the
opening of a pseudogap \cite{Kataev92,Wzietek96,Kanoda97}. The
softening of ultrasound modes exhibiting
pronounced minima at 38\thinspace K and 46\thinspace K for $\kappa $-H$_{8}$%
-Br and $\kappa $-Cu(NCS)$_{2}$, respectively, was attributed to the same
effect \cite{Frikach00}.\newline
Besides these anomalies around 50\thinspace K, unusual time dependencies in
magnetic and transport properties have been reported for both deuterated and
hydrogenated $\kappa $-Br near 80\thinspace K \cite%
{Kawamoto97,Su98a,Su98b,Su98c}. For $\kappa $-H$_{8}$-Br, the
superconducting properties have been found to depend on the
thermal history, in particular on how fast the sample had been
cooled through 77\thinspace K. As mentioned above, the ground
state of $\kappa $-D$_{8}$-Br is strongly sample-dependent: both
superconducting as well as non-superconducting crystals are found.
Furthermore, superconducting as well as insulating
(possibly antiferromagnetic) phases in separated volume parts of the {\em %
same} sample have been reported. Their relative volume fraction was found to
depend on the cooling rate $q_{c}$ employed at around 80\thinspace K \cite%
{Kawamoto97,Ito00}: in quenched-cooled crystals ($q_{c}\sim
-100$\thinspace K/min), a strong decrease of the diamagnetic
signal has been observed, which has been interpreted as indicating
a suppression of the superconducting in favour of the magnetic
phase.

Concerning both the pairing mechanism and the symmetry of the
superconducting order parameter, the experimental situation is
still unsettled: a number of recent experiments provide strong
arguments for an isotropic gap structure and indicate the
particular role of lattice degrees of freedom in the pairing
interaction \cite{Kini96,Pintschovius97,Elsinger00}. This
conflicts sharply with the results of other experiments, notably
NMR measurements performed in finite magnetic fields
\cite{DeSoto95,Mayaffre95,Kanoda96}, that reveal low energy
excitations indicative of an anisotropic pairing state with nodes
in the gap. For a review on the controversy on the superconducting
state, see for example \cite{Lang96,Wosnitza99}. In connection
with the above controversy on the state below $T_{c},$ questions
arise about the origin of the anomalous properties at $T>T_{c}$
and their interrelation to the superconducting state. In addition,
one may ask to what extent a comparison can be drawn to the
high-$T_c$ cuprates and - an obvious concern for the present
molecular systems - what the role of lattice degrees of freedom is
for the above-mentioned anomalies.

A most sensitive tool, to probe the lattice properties and their coupling to
the electronic degrees of freedom, is provided by measurements of the linear
thermal expansion coefficient $\alpha (T)=\partial \ln l(T)/\partial T$,
where $l(T)$ is the sample length. To this end we have initiated a
systematic thermal expansion study on the title compounds. In recent
comparative thermal expansion experiments focusing on the uniaxial-pressure
coefficients of $T_{c}$, $\partial T_{c}/\partial p_{i}$, for various (ET)$%
_{2}$X compounds, we found that the {\em interlayer}-pressure
effect is of crucial importance for $T_{c}$
\cite{Mueller00,Mueller01}. We present here the extension of that
work to higher temperatures covering also the unusual normal-state
properties. Our thorough comparative studies have enabled us to
distinguish three different kinds of anomalies in $\alpha (T)$ in
the temperature range $1.5\,{\rm K}\leq T\leq 80$\thinspace K.
These are (A) the phase transitions at $T_{c}$ or $T_{N}$
determining the actual ground state of the system, (B)
well-defined $\alpha (T)$ maxima at intermediate temperatures
around $30\,{\rm K}-50\,{\rm K}$ for the superconducting salts and
(C) second-order phase-transition-like jumps in $\alpha (T)$ due
to a
kinetic, glass-like transition at higher temperatures around $70\,{\rm K}%
-80\,{\rm K}$. As will be shown below, the latter originate in
relaxation effects of the ethylene endgroups of the ET molecules.
Such a glass-like freezing process of the terminal ethylene
moieties provides a natural explanation for the above-mentioned
time-dependent transport and magnetic properties found for $\kappa
$-Br. We will also discuss possible implications of disorder for
the ground-state properties. Furthermore, our results hint at a
phase transition, i.e.\ a real gap formation that involves only a
minor part of the Fermi surface, in the superconducting salts
$\kappa$-Br and $\kappa$-Cu(NCS)$_2$ at $T^{\ast}=38$\,K and
45\,K, respectively, as opposed to a pseudogap on the whole Fermi
surface. Concerning the ground-state properties, we will present
results on the anisotropy of the uniaxial-pressure coefficients of
$T_{c}$ for $\kappa $-Br that supplement our previous studies
\cite{Mueller00,Mueller01} and give clear thermodynamic evidence
for a second-order phase transition at $T_{N}$ for $\kappa$-Cl.

\section{Experiment}\label{Experiment}
The coefficient of thermal expansion has been measured utilizing a
high-resolution capacitance dilatometer \cite{Pott83} with a
maximum sensitivity corresponding to $\Delta l/l=10^{-10}$. Length changes $%
\Delta l(T)=l(T)-l(T_{0})$, where $T_{0}$ is the base temperature,
were detected upon both heating and cooling the sample. The
coefficient of thermal expansion $\alpha (T)=\frac{1}{l(T)}\cdot
\frac{\partial l(T)}{\partial T}$ is approximated by the
differential quotient $\alpha (T)\approx \frac{\Delta
l(T_{2})-\Delta l(T_{1})}{l(300{\rm K)\cdot (T_{2}-T_{1})}}$ with $%
T=(T_{1}+T_{2})/2$.\newline
The single crystals used were synthesized by the standard
electrocrystallization technique with typical dimensions of $(0.5-1.5)$%
\thinspace mm perpendicular to the highly conducting planes and $(0.5-2.5)$%
\thinspace mm along both in-plane axes. In the present work ten crystals
have been studied: one $\kappa $-Cl sample ($T_{N}=27.8$\thinspace K), four
crystals of $\kappa $-Br and five of $\kappa $-Cu(NCS)$_{2}$ including
samples with various isotope substitutions. Due to the particular shape of
the $\kappa $-Cl and $\kappa $-Br crystals used \cite{Wang90}, it was not
possible to measure $\alpha (T)$ along all three principal axes on the {\em %
same} sample. For $\kappa $-H$_{8}$-Br measurements perpendicular
to the planes ($b$-axis) and along one not specified in-plane axis
were performed on crystal \#\thinspace 2 ($T_{c}=11.8$\thinspace
K), while measurements along the in-plane $a$- and $c$-axis were
performed on crystal \#\thinspace 3 ($T_{c}=11.5$\thinspace K).
The excellent quality of both crystals is reflected by both the
high transition temperatures and relatively small transition
widths of $\Delta T_{c}=500$\,mK, as determined by a thermodynamic
measurement as well as the pronounced anisotropy in the uniaxial
thermal expansion coefficients. To check for reproducibility, we
measured another crystal of $\kappa $-H$_{8}$-Br, \#\thinspace 1,
along the out-of-plane axis and found (besides a slightly lower
$T_{c}$ value of 11.5\thinspace K) almost identical expansivity
behavior to that for crystal \#\thinspace 2 in the whole
temperature range investigated, i.e.\ up to 200\thinspace K. The
crystals \#\thinspace 1 and \#\thinspace 2 have also
been used in our previous study on the uniaxial-pressure coefficients of $%
T_{c}$ \cite{Mueller01}. For a deuterated crystal, $\kappa
$-D$_{8}$-Br, which was measured along the $b$-axis, no
superconducting phase-transition anomaly was found. Regarding the
above-mentioned cooling-rate dependence of the ground state, we
note that for all samples investigated no changes of the {\em
low-temperature} expansivity behavior were observed within the
parameters accessible with our experimental setup, allowing cooling rates $%
q_{c}$ ranging from $-1\,{\rm K/h}$ up to $-300\,{\rm K/h}$. For $\kappa $%
-Cu(NCS)$_{2}$, measurements along the out-of-plane $a^{\ast }$-axis were
performed on two crystals of the pure compound ($T_{c}=9.2$\thinspace K) and
three crystals with isotope substitutions either on the ET or anion sites.
In each case, a pronounced and relatively sharp phase-transition anomaly was
found at $T_{c}$. For comparison, we studied deuterated $\kappa $-(D$_{8}$-ET)%
$_{2}$Cu(NCS)$_{2}$ ($T_{c}=9.95$\thinspace K), anion-labeled $\kappa $-(ET)$%
_{2}$Cu($^{15}$N$^{13}$CS)$_{2}$ ($T_{c}=9.3$\thinspace K) and $\kappa $-(D$%
_{8}$$^{13}$C$_{4}$$^{34}$S$_{8}$-ET)$_{2}$Cu(NCS)$_{2}$ ($T_{c}=9.8$%
\thinspace K) where eight sulfur atoms of the ET molecule have been labeled
with $^{34}$S, the four ethylene carbon atoms with $^{13}$C and the eight
hydrogen atoms with deuterium. As pointed out in \cite{Kini96}, the latter
compound shows a 'normal' BCS-like mass isotope effect on $T_{c}$ for the $%
^{13}$C$^{34}$S substitution and an 'inverse' isotope effect upon
replacing the hydrogen atoms of the ethylene endgroups by
deuterium; no isotope effect on $T_{c}$ is observed for the
anion-labeled salt \cite{Kini96b}. In the present paper we refrain
from comparing superconducting properties for the afore-mentioned
$\kappa $-Cu(NCS)$_{2}$ samples containing isotope substitutions,
but have tended to focus on the effect of isotope substitution on
the expansivity anomalies at intermediate temperatures (B). For
the deuterated sample, the in-plane $b$- and $c$-axis thermal
expansion coefficients have also been measured. This crystal of
excellent quality is identical to that studied in
\cite{Mueller00}. For the determination of $T_{c}$ and $T_{N}$, we
use the standard "equal-areas" construction in a plot $\alpha
(T)/T$ {\em vs} $T$. A small misalignment of the crystal
orientation of about $\pm 5^{\circ}$ from the exact alignment
cannot be excluded.

\section{Classification of thermal expansion anomalies}
\label{Aufhaenger} 
\begin{figure}[tbp]
\centerline{\psfig{figure=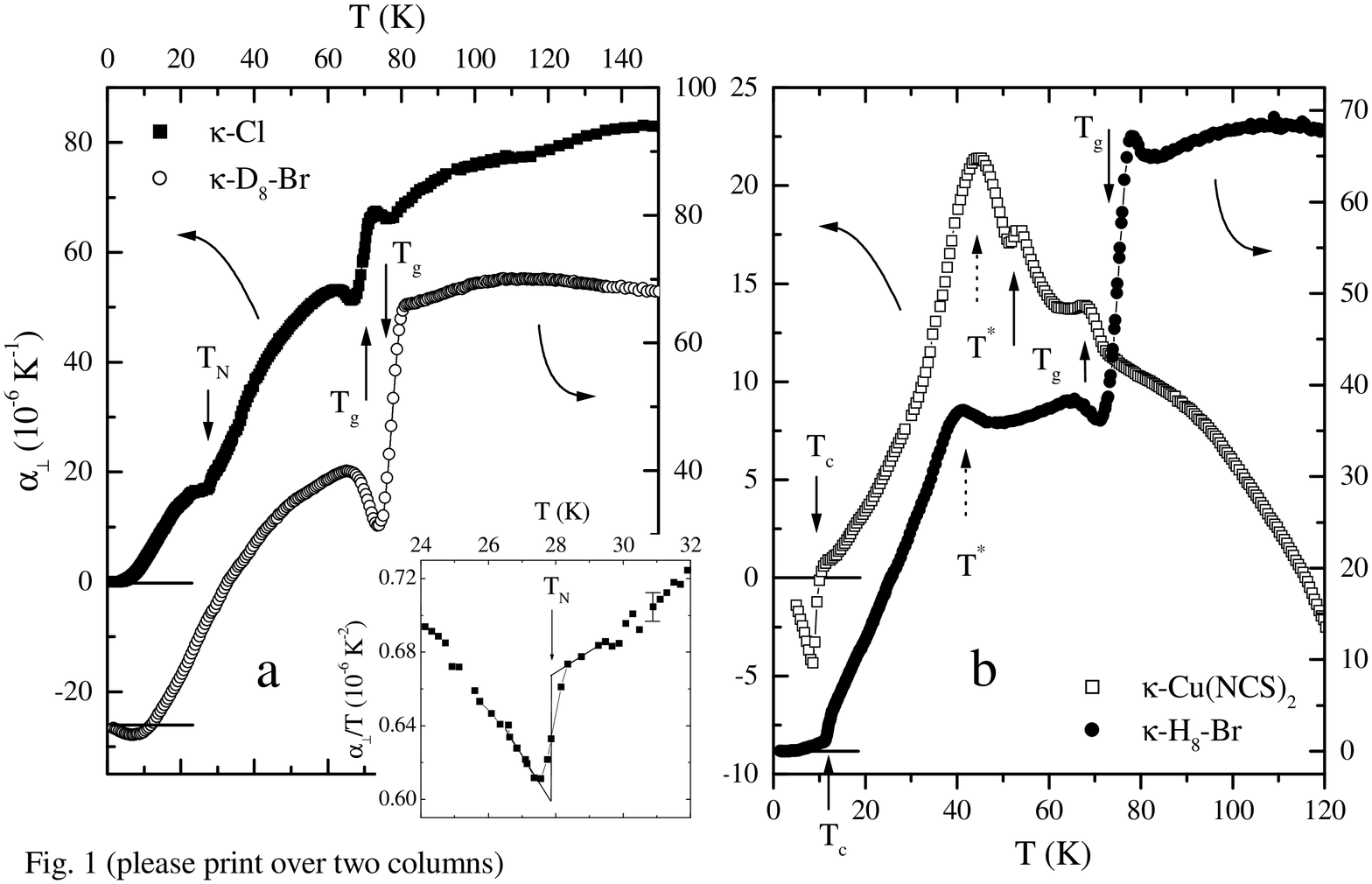,width=\textwidth}}
\vspace{0.5cm}
\caption{Linear thermal expansion coefficient perpendicular to the planes, $%
\protect\alpha_{\perp}$, {\em vs} $T$ for various salts $\protect\kappa$-(ET)%
$_2$X. (a) non-superconducting $\protect\kappa$-Cl and $\protect\kappa$-D$_8$%
-Br, (b) superconducting $\protect\kappa$-H$_8$-Br and $\protect\kappa$%
-Cu(NCS)$_2$. For clarity, different scales have been used along
the
ordinates. The inset shows details of $\protect\alpha_{\perp}$ for X=Cu[N(CN)%
$_2$]Cl as $\protect\alpha_{\perp} / T$ {\em vs} $T$. Arrows indicate
different kinds of anomalies as explained in the text.}
\label{motivation}
\end{figure}

Fig.\ \ref{motivation} shows the thermal
expansion coefficient measured perpendicular to the highly
conducting planes, $\alpha_{\perp }$, for $\kappa $-(ET)$_{2}$X
with X=Cu[N(CN)$_{2}$]Cl, X=Cu[N(CN)$_{2}$]Br and X=Cu(NCS)$_{2}$
over an extended temperature range. Fig.\ \ref{motivation}(a)
compares $\alpha _{\perp }(T)$ of the non-superconducting salts
$\kappa $-Cl and $\kappa $-D$_{8}$-Br. In Fig.\
\ref{motivation}(b) we show the expansivity data of
superconducting $\kappa $-H$_{8}$-Br and $\kappa $-Cu(NCS)$_{2}$.
Note the different scales of the ordinate for each case. For the
various compounds, a number of anomalies is observed for
temperatures $T\leq 80$\thinspace K as indicated by the arrows. As
will be discussed in more detail below, three different kinds of
anomalies can be distinguished that correspond to different
temperature ranges:

At temperatures around $70\,{\rm K}-80\,{\rm K}$ (C), large second-order
phase-transition-like anomalies are found for $\kappa $-Cl and both kinds of
$\kappa $-Br salts. Their characteristic temperatures are labeled as $T_{g}$%
. For $\kappa $-Cu(NCS)$_{2}$, a smaller but distinct step-like anomaly at
70\thinspace K and a second one at around 53\thinspace K can be observed. As
will be shown below, these anomalies are of the same origin, namely due to a
kinetic, glass-like transition where, below $T_{g}$, a certain disorder in
the positional degrees of freedom of the [(CH$_{2}$)$_{2}$] endgroups of the
ET molecules becomes frozen in.

In the intermediate temperature range around $40\,{\rm K}-50\,{\rm
K}$ (B) both superconducting compounds exhibit a pronounced local
maximum in $\alpha (T)$ at a temperature labeled as $T^{\ast }$,
Fig.\ \ref{motivation}(b), while the features are absent in both
non-superconducting salts, see Fig.\ \ref{motivation}(a). These
anomalies have been reproduced in detail for all five crystals of
$\kappa $-Cu(NCS)$_{2,}$ and both of the crystals of $\kappa
$-H$_{8}$-Br measured (see section \ref{Experiment}). They can
thus be regarded as intrinsic properties. Most pronounced is the
sharp maximum in $\alpha _{\perp }(T)$ for the $\kappa
$-Cu(NCS)$_{2}$ compound at 45\thinspace K indicating a
significant change in the temperature dependence of the interlayer
lattice parameter at this temperature, i.e.\ an additional strong
contraction of the interlayer lattice parameter $a^{\ast }$ upon
cooling through $T^{\ast }$. While the overall expansivities of
$\kappa $-Cl and $\kappa $-Br show the usual increase in $\alpha $
as $T$ increases, for $\kappa $-Cu(NCS)$_{2}$ an anomalous
temperature dependence is revealed for temperatures $T>T^{\ast }$,
i.e., $\alpha _{\perp }(T)$ decreases with increasing temperature.
Above $T\approx 120$\,K, it even becomes negative corresponding to
a progressive reduction of the interlayer distance upon warming.
Then, at $T\approx 175$\thinspace K, $\alpha _{\perp }(T)$ passes
through a broad minimum and becomes positive again above
220\thinspace K (not shown). The latter findings have been pointed
out already in a previous thermal expansion study \cite{Kund95}.
In the following, we concentrate on the anomalies at $T^{\ast}$
found for the superconducting salts. Subtracting the interpolated
lattice background and judging from the shape of the separated
anomalies, i.e.\ their sharpness and magnitude, it is tempting to
attribute these effects to a true phase transition. We will
discuss its relation to the various anomalies reported for these
compounds in the literature and propose a generalized scenario for
the $\kappa$-(ET)$_2$X title compounds in terms of two competing
order parameters associated with $T_c$ and $T^{\ast}$, forming on
disjunct parts of the Fermi surface.

At low temperatures, i.e.\ in the temperature range (A), the
thermal expansion behavior is dominated by the phase-transition
anomalies characterizing the respective ground state of the
system: the ambient-pressure superconductors $\kappa $-H$_{8}$-Br
and $\kappa $-Cu(NCS)$_{2}$ show pronounced negative second-order
phase-transition anomalies in $\alpha _{\perp}(T)$ at
$T_{c}=11.8$\thinspace K and 9.3\thinspace K, respectively. Since
the discontinuities in $\alpha _{i}$, $\Delta \alpha _{i}$, at
$T^{\star }$, typical for a second-order phase transition, are
related via the Ehrenfest relation to the uniaxial-pressure
dependencies of $T^{\star}$ along the $i$-axis in the limit of
vanishing pressure:
\begin{equation}
\left( \frac{\partial T^{\star }}{\partial p_{i}}\right) _{p_{i}\rightarrow
0}=V_{{\rm mol}}\cdot T^{\star }\cdot \frac{\Delta \alpha _{i}}{\Delta C},
\label{eq:ehrenfest}
\end{equation}%
where $V_{{\rm mol}}$ denotes the molar volume and $\Delta C$ the
discontinuity in the specific heat at $T^{\star }$, the large
negative discontinuities in $\alpha _{\perp}(T)$ at $T_{c}$
demonstrate that uniaxial stress perpendicular to the planes
causes a strong suppression of $T_{c}$. In section
\ref{normalstate} we will discuss the anisotropy of the lattice
response at $T_{c}$ for the superconducting $\kappa $-H$_{8}$-Br
and $\kappa $-Cu(NCS)$_{2}$ salts.\\ The inset of Fig.\
\ref{motivation}(a) gives details of $\alpha_{\perp}(T)$ for
$\kappa $-(ET)$_{2}$Cu[N(CN)$_{2}$]Cl in a representation $\alpha
/T$ {\em vs} $T$. A distinct negative jump reminiscent of a
second-order phase transition can be resolved at 27.8\thinspace K.
This is about the same temperature at which $^{1}$H-NMR and
magnetization measurements revealed the onset of 3D
antiferromagnetic order \cite{Miyagawa95}. Our data thus provide,
for the first time in a thermodynamic quantity beyond
magnetization, clear evidence for a second-order phase transition
in $\kappa$-(ET)$_{2}$Cu[N(CN)$_{2}$]Cl at $T_{N}$.

After analyzing the glassy anomalies at higher temperatures (C) in the
following section \ref{Glas} we will focus on the expansivity properties in
the low- (A) and intermediate-temperature range (B) in sections \ref{AFI}
and \ref{normalstate}. A conclusion is given in section \ref{conclusion}.

\section{The glass-like transition} \label{Glas}
In positionally and/or orientationally disordered systems,
relaxation processes can lead to glass-like transitions where,
below a certain temperature $T_{g}$, a short-range order
characteristic for this temperature is frozen in. These relaxation
phenomena are not limited to ''classical'' glass-forming materials
like undercooled liquids: corresponding orientational degrees of
freedom can lead to glass-like transitions also in polymeric or
crystalline materials with large organic molecules as, e.g.,
single-crystalline C$_{60}$ \cite{Gugenberger92}. Such transitions
occur when the (structural) relaxation time of the relevant
molecular movements, which is growing exponentially with
decreasing temperature, exceeds the characteristic time scale of
the experiment. Then the system cannot reach its equilibrium state
before the temperature further decreases. As a result, the
relevant degrees of freedom cannot be thermally excited and, thus,
no longer contribute to quantities such as the specific heat:
$C^{{\rm glass}}(T)=0$ for $T<T_{g}$ and $C^{{\rm glass}}(T)>0$
for $T>T_{g}$ with $C^{{\rm glass}}$ representing the additional
contribution to the heat capacity. As the volume thermal expansion
coefficient $\beta (T)=\sum_{i}\alpha _{i}(T)$, with $i=a,b,c$, is
related to the specific heat via the Gr\"uneisen relation
\begin{equation}
\beta (T)=\Gamma \cdot \frac{\kappa _{T}}{V_{{\rm mol}}}\cdot
C_{V}(T), \label{eq:Gruneisen}
\end{equation}%
where $\kappa _{T}$ denotes the isothermal compressibility and
$\Gamma $ a generalized  Gr\"uneisen parameter, both methods are
very well suited for studying glassy phenomena. In particular, the
thermal expansion coefficient is extremely sensitive to structural
rearrangements and allows for studying the anisotropy of the
glassy effects. Glass-like transitions can be identified by the
following characteristics: (i) a step-like increase of $C_{V}$ and
$\beta $ upon heating through $T_{g}$, (ii) the occurrence of
hysteresis around $T_{g}$ and (iii) a cooling-rate dependent
$T_{g}$ value \cite{Cahn91}. Due to the general nature of the
molecular motions involved in the relevant relaxation processes,
the phenomenology of glass transitions as observed in undercooled
liquids applies in certain limits also to the glass-forming
subsystems in polymeric or crystalline materials \cite{Angell95}
like in the quasi-2D organic superconductors studied in this work.

In a recent ac-calorimetry study of the $\kappa$-Br and the
$\kappa$-Cl, salt glass transitions have been reported for both
compounds \cite{Saito99,Akutsu00}. The authors observed step-like
anomalies in the heat capacity around 100\,K. The activation
energy for the relevant relaxation process has been deduced by
analyzing the frequency dependence of the glass-transition
temperature. The transition was attributed to a freezing out of
the intramolecular motion of the ethylene endgroups of the ET
molecules. It was claimed that an extrapolation of the
frequency-dependent glass-transition temperatures to frequencies
corresponding to "the time scale of our daily life" ($\sim
10^3$\,s) would give a $T_g$ value of about 80\,K. At about the
same temperature, Kund {\em et al.} reported an anomalous thermal
expansion behavior of the $\kappa$-Br and $\kappa$-Cl salts
\cite{Kund93,Kund94}: these authors found abrupt changes of
$\alpha_i(T)$ at 80\,K and 73\,K, respectively, which they
attributed to second-order phase transitions. An explanation was
given in terms of an order-disorder phase transition of the
terminal ethylenes. The high-resolution thermal expansion results
presented here provide convincing evidence for a glass-like
transition in the $\kappa$-Br and $\kappa$-Cl salts. This
confirms, on the one hand, the above-mentioned specific heat
results \cite{Saito99,Akutsu00} and clarifies on the other, the
nature of the thermal expansion anomalies previously reported by
Kund {\em et al.} \cite{Kund93,Kund94}. For the first time, we
report a glass-like transition also for
$\kappa$-(ET)$_2$Cu(NCS)$_2$. A careful study of the cooling-rate
dependence of $T_g$ enables us to evaluate
the activation energy for the relevant relaxation process. For the $\kappa$%
-Br compound, we discuss the effect of H-D isotope substitution in the
terminal ethylenes. %The anisotropy of the thermal expansion anomalies will
%be shown to reflect the peculiarities of the crystal structures of
%the present systems.
We further discuss the systematics of the glassy behavior for the title
compounds and the relation to various time dependencies reported in
literature including possible implications for the low-temperature
properties.

\subsection{Phenomena and analysis}\label{glas1}
\begin{figure}[tbp]
\centerline{\psfig{figure=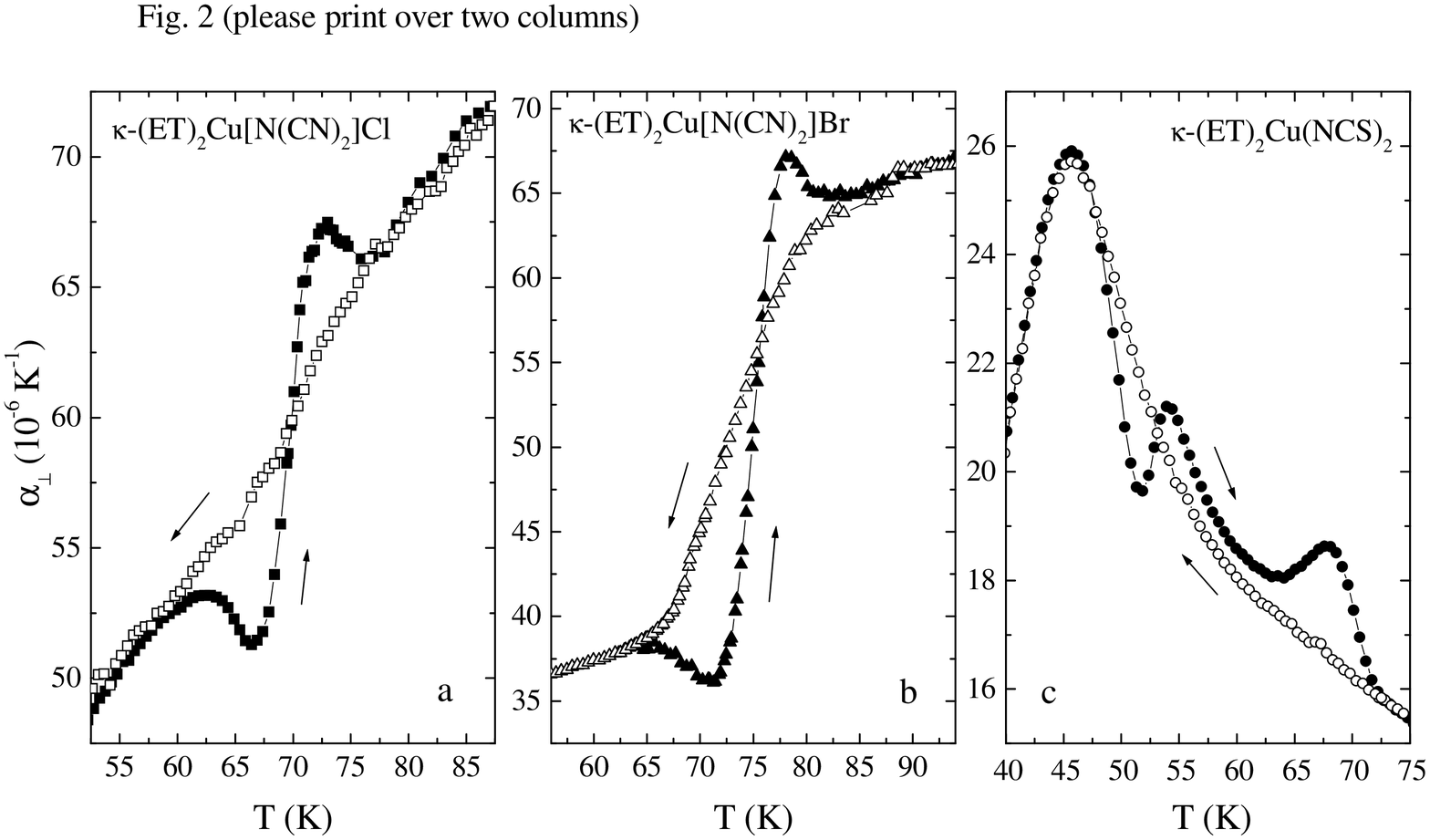,width=\textwidth}}
\vspace{0.5cm}
\caption{Coefficient of thermal expansion perpendicular to the planes, $%
\protect\alpha_{\perp}$, {\em vs} $T$ close to the glass-like
transition for
(a) $\protect\kappa$-(ET)$_2$Cu[N(CN)$_2$]Cl, (b) $\protect\kappa$-(ET)$_2$%
Cu[N(CN)$_2$]Br and (c) $\protect\kappa$-(ET)$_2$Cu(NCS)$_2$. Closed and
open symbols denote heating and cooling curves, respectively.}
\label{glass1}
\end{figure}

Fig.\ \ref{glass1} shows on expanded scales the
anomalies in $\alpha _{\perp }(T)$ in the temperature range (C) for the $%
\kappa $-Cl, $\kappa $-H$_{8}$-Br and $\kappa $-Cu(NCS)$_{2}$ salts. A
distinct hysteresis is seen between the heating (closed symbols) and cooling
curves (open symbols). In the heating curves, the former two salts reveal
pronounced jump-like anomalies at $T_{g}\approx 70\,$K ($\kappa $-Cl) and $%
75 $\thinspace K ($\kappa $-Br) with characteristic under- and
overshoots at the low- and high-temperature sides of the $\alpha
(T)$ discontinuity, respectively. While the heating and cooling
data coincide further outside the transition region, they differ
markedly close to $T_{g}$: upon cooling no under- and overshoot
behavior can be observed and the $\alpha_{\perp}(T)$ curves are
reminiscent of broadened second-order phase-transition anomalies.
Such a hysteretic behavior is not expected for thermodynamic phase
transitions but, as mentioned above, it is characteristic of
glass-like transitions, where instead of a well-defined transition
temperature, one meets a ''glass-transformation temperature
interval'' \cite{Cahn91,deBolt76}. In contrast to the
isostructural $\kappa $-Cl and $\kappa $-Br salts where a single
anomaly occurs, the heating curve for the $\kappa $-Cu(NCS)$_{2}$
compound shows a sequence of two somewhat smaller step-like
anomalies at around 70\thinspace K and 53\thinspace K, cf.\ Fig.\
\ref{glass1}(c). The hysteretic behavior indicates that both
features are glassy in nature, with the transition region
extending over a wide range from about 48\thinspace K up to
73\thinspace K.\newline We note that, besides the actual value of
$T_{g}$, which will be discussed below, the details of the heating
curves, especially the characteristics of the under- and overshoot
behavior, also depend on the thermal history: for all three
systems, the overshoot on reheating was found to be more
pronounced when the heating rate $|q_{h}|$ was larger than the
preceding cooling rate $|q_{c}|$. On the other hand, if
$|q_{h}|<|q_{c}|$, there was relatively little overshoot above
$T_{g}$ but undershoot for $T<T_{g}$. This
is precisely what one expects for ''classical'' glass-forming systems \cite%
{Cahn91}.\newline The question at hand is which molecular motions
are involved in the relaxation process, i.e., what kind of
structural degrees of freedom are frozen in at low temperatures.
To this end it is helpful to evaluate the activation barrier for
the relevant relaxation process from the cooling-rate ($q_{c}$)
dependence of $T_{g}$ \cite{deBolt76}.

\subsubsection{$\protect\kappa$-(ET)$_2$Cu[N(CN)$_2$]Br and $\protect\kappa$%
-(ET)$_2$Cu[N(CN)$_2$]Cl}
\begin{figure}[tbp]
\centerline{\psfig{figure=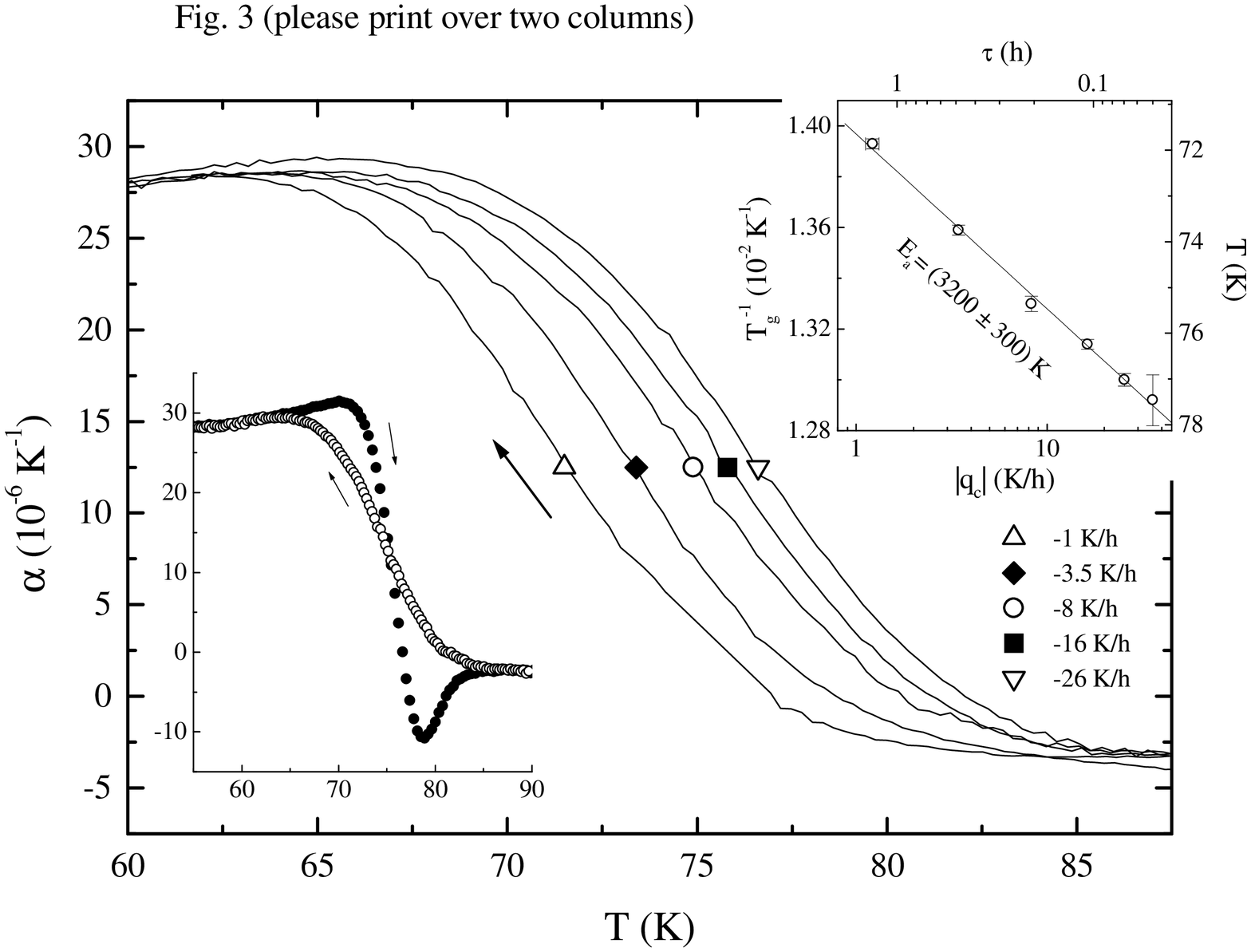,width=\textwidth}}
\vspace{0.5cm} \caption{Linear thermal expansion coefficient,
$\protect\alpha$, {\em vs} $T$
measured along one non-specified in-plane axis of $\protect\kappa$-(ET)$_2$%
Cu[N(CN)$_2$]Br in the vicinity of the glass transition for
varying cooling rates $q_c$. The definition of the
glass-transition temperature $T_{{\rm g}}$ is given in the text.
Insets: hysteresis between heating and cooling curves
around $T_g$ (left side) and Arrhenius plot of $T_{{\rm g}}^{-1}$ {\em vs} $%
|q_c|$ and $\protect\tau$ (right side).}
\label{glass2}
\end{figure}

\begin{figure}[tbp]
\centerline{\psfig{figure=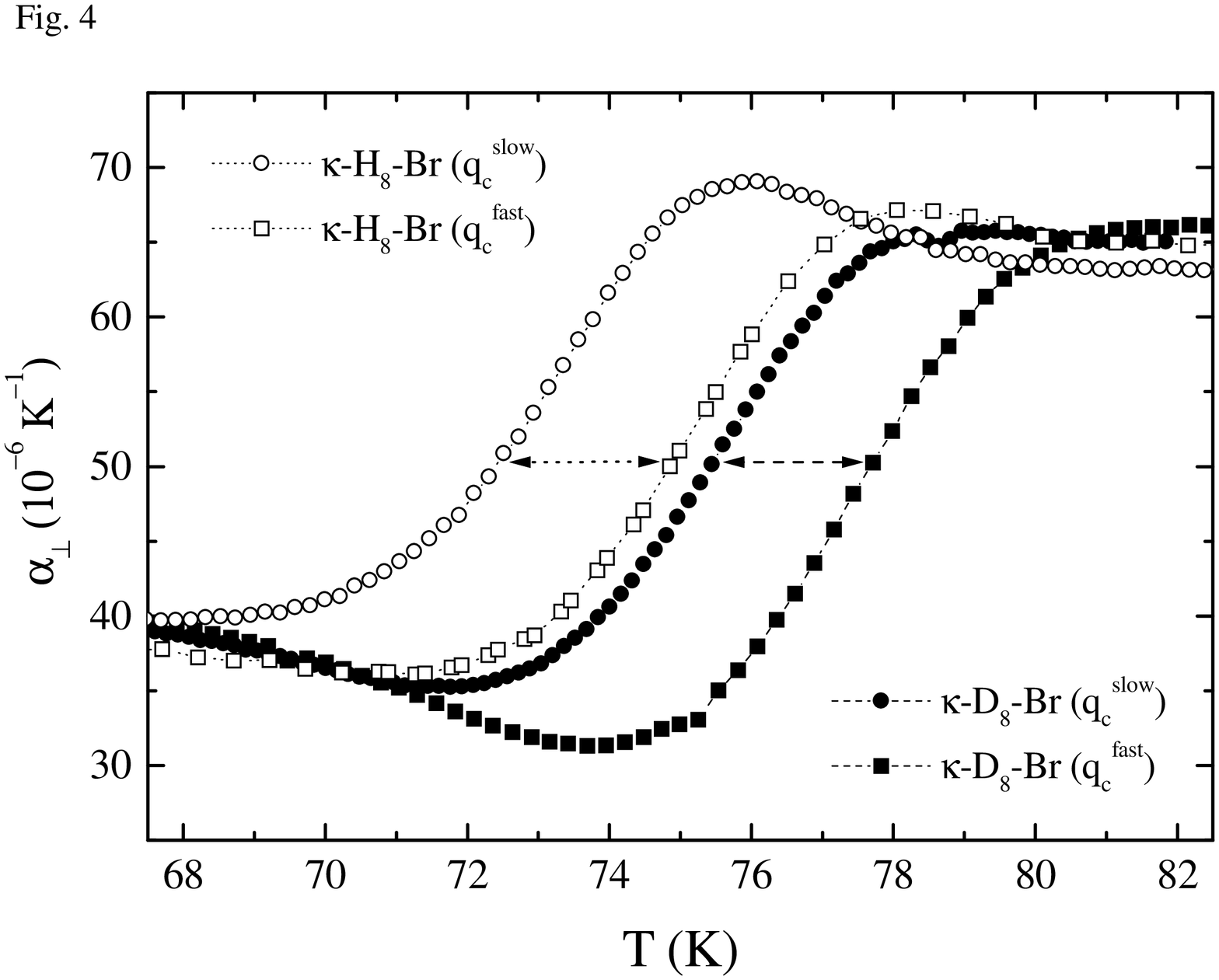,width=\textwidth}}
\vspace{0.5cm} \caption{Coefficient of thermal expansion
perpendicular to the planes, $\protect\alpha_{\perp}$, {\em vs}
$T$ of hydrogenated (open symbols) and deuterated (closed symbols)
$\protect\kappa$-(ET)$_2$Cu[N(CN)$_2$]Br. Data were taken upon
heating after cooling the crystals with two different rates
$q_c^{{\rm slow}}$ and $q_c^{{\rm fast}}$ (see text) which results
in a lower and upper value of $T_g$ for the slow and fast cooling
history, respectively, in each case, as indicated by the arrows.}
\label{glass3}
\end{figure}

Fig.\ \ref{glass2} shows the linear thermal expansion coefficient $%
\alpha (T)$ measured along one non-specified in-plane axis of the $\kappa $%
-Br salt at different cooling rates $q_{c}$. The inset on the left
side illustrates the hysteretic behavior around $T_{g}$. For the
transition temperature $T_{g}(q_{c})$, we use the midpoints of the
somewhat broadened step-like anomalies in the respective cooling
curves. The figure clearly shows that the transition shifts to
higher temperatures with increasing cooling rate $|q_{c}|$,
whereas the shape of the curves remains unchanged. This behavior
is well understood for a glass transition: cooling or heating at a
continuous rate, $q=dT/dt$, may be thought of as a sequence of
differential temperature steps, $\Delta T$, interrupted by
intervals of duration $\Delta t=\Delta T/q$ at which $T={\rm
const}$. The system remains in equilibrium as long as $\Delta t$
is much longer than the relaxation time $\tau $ which, for
structural rearrangement processes in solids, is known to grow
exponentially upon lowering the temperature \cite{Cahn91}. At high
temperatures, $\tau $ is small and thus, $\tau \ll \Delta t$. As soon as $%
\tau \approx \Delta t$ upon cooling, the relaxation of one step is not
completed before the temperature further decreases. %near $T_g$,
%$\tau \approx dt$, so relaxation of one step is not complete
%before the temperature further decreases.
The greater the cooling rate, the less time remains for
relaxation; since, upon increasing $|q_c|$, $\Delta t\sim 1/|q_c|$
becomes smaller, the transition defined by $\tau \approx \Delta t$
occurs at higher temperatures \cite{Cahn91}. The inset on the
right side of Fig.\ \ref{glass2} shows an Arrhenius plot of the
inverse of the so-derived glass-transition temperatures,
$T_{g}^{-1}$, {\em vs} $|q_{c}|$. The data nicely follow a linear
behavior as expected for a thermally activated relaxation time
\cite{deBolt76,Nagel00}
\begin{equation}
\tau (T)=\nu_{0}^{-1}\cdot e^{E_{{\rm a}}/k_{B}T}, \label{eq:tau}
\end{equation}
where $E_{{\rm a}}$ denotes the activation energy barrier and
$k_{B}$ the Boltzmann constant. The pre\-factor represents an
attempt frequency $\nu_{0}$.\newline Based on a simple two-state
model \cite{comment1a}, Nagel {\em et al.} have quantified the
above-sketched idea of relating the cooling rate with the
relaxation time \cite{Nagel00}. They derived a criterion for the
glass-transition temperature that allows for a determination of
$\nu_{0}$ and $\tau $: $\tau (T_{{\rm
g}})=\frac{T_{g}}{|q_c|}\cdot \frac{k_{B}T_{g}}{E_{{\rm a}}}$. In
a first approximation we set $\tau (T_{g})\sim \frac{1}{|q_c|}$
yielding: $\ln |q_c|=-E_{{\rm a}}/k_{B}T_{g}+{\rm const}$. This is
in excellent agreement with the observed cooling-rate dependence
of $T_{g}$, cf.\ right inset of Fig.\ \ref{glass2}. A linear fit
to the data of Fig.\ \ref{glass2} yields $E_{{\rm a}}=(3200\pm
300)$\thinspace K and $\nu_{0}=5 \times 10^{15 \pm 1.5}\,{\rm
Hz}$.\newline As mentioned above, there are
positional/orientational degrees of freedom for the
[(CH$_{2}$)$_{2}$] endgroups which represent the most deformable
parts of the ET molecules \cite{Rahal97}. For $\kappa$-Br, the
ethylene groups are known to be partially disordered at room
temperature and to become ordered in the eclipsed conformation
below 127\thinspace K \cite{Geiser91}. The characteristic
activation energy of the [(CH$_{2}$)$_{2}$] conformational motion
was determined as $E_{{\rm a}}=2650$\,K by $^{1}$H-NMR
measurements \cite{Wzietek96}. This value has to be compared to
2400\thinspace K as estimated from ac calorimetry \cite{Akutsu00},
$(2000\pm 500)$\thinspace K from resistance measurements of
structural relaxation kinetics \cite{Tanatar99} and 2600\thinspace
K also from resistance-relaxation measurements \cite{Su98b}. The
similar size of the activation energy derived here, compared to
the above numbers, suggests that the [(CH$_{2}$)$_{2}$] endgroups
are the relevant entities for the relaxation process observed in
the thermal expansion. A direct check for this possibility is
provided by measuring the mass-isotope effect on $T_{g}$. For a
deuterated $\kappa $-Br salt due to the higher mass of its
[(CD$_{2}$)$_{2}$] units, a shift of $T_{g}$ is expected: for
$\kappa $-D$_{8}$-Br the relaxation time $\tau $ of the terminal
(CD$_{2}$)$_{2}$ groups at a given temperature should exceed that
for the $\kappa $-H$_{8}$-Br compound at the same temperature.
Using the above criterion for $T_{g}$, a longer relaxation time
means that for a given cooling rate the system falls out of
equilibrium at higher temperatures. Fig.\ \ref{glass3} compares
$\alpha _{\perp }(T)$ for the hydrogenated compound, $\kappa
$-H$_{8}$-Br (open symbols), with that for the deuterated one,
$\kappa $-D$_{8}$-Br (closed symbols). For both compounds, two
heating curves are shown taken after slow and fast
cooling history ($q_{c}^{{\rm slow}}\approx -25$\thinspace K/h and $q_{c}^{%
{\rm fast}}\approx -150$\thinspace K/h). The figure clearly demonstrates
that the $T_{g}$ values for the deuterated $\kappa $-Br salt are shifted by
about 3\thinspace K to higher temperatures compared to that for $\kappa $-H$%
_{8}$-Br. To check for reproducibility, we measured a second
sample of the hydrogenated salt from a different batch. We found
that under similar conditions, the $T_{g}$ value was reproduced
within about 250\,mK.
%This result confirms a true
%shift of the glass-transition temperature due to isotope
%substitution in the ethylene system of
%$\kappa$-(ET)$_2$Cu[N(CN)$_2$]Br.
If one simply assumes that the ratio of the relaxation times for
the terminal [(CD$_{2}$)$_{2}$] and [(CH$_{2}$)$_{2}$] units scale
with the square root of their mass ratios one would expect a
positive shift of $T_{g}$ of 3.4\thinspace \% upon $^{1}{\rm
H}\rightarrow ^{2}$D substitution (here we consider the activation
barrier $E_{{\rm a}}$ to be identical for both isotopes). Thus the
observation of a $^{1}$H-$^{2}$D isotope effect on $T_{g}(q_{c})$
of about 4\thinspace \% provides clear evidence for a relaxation
of the ethylene moieties as the origin of the observed glassy
phenomena.

\begin{figure}[tbp]
\centerline{\psfig{figure=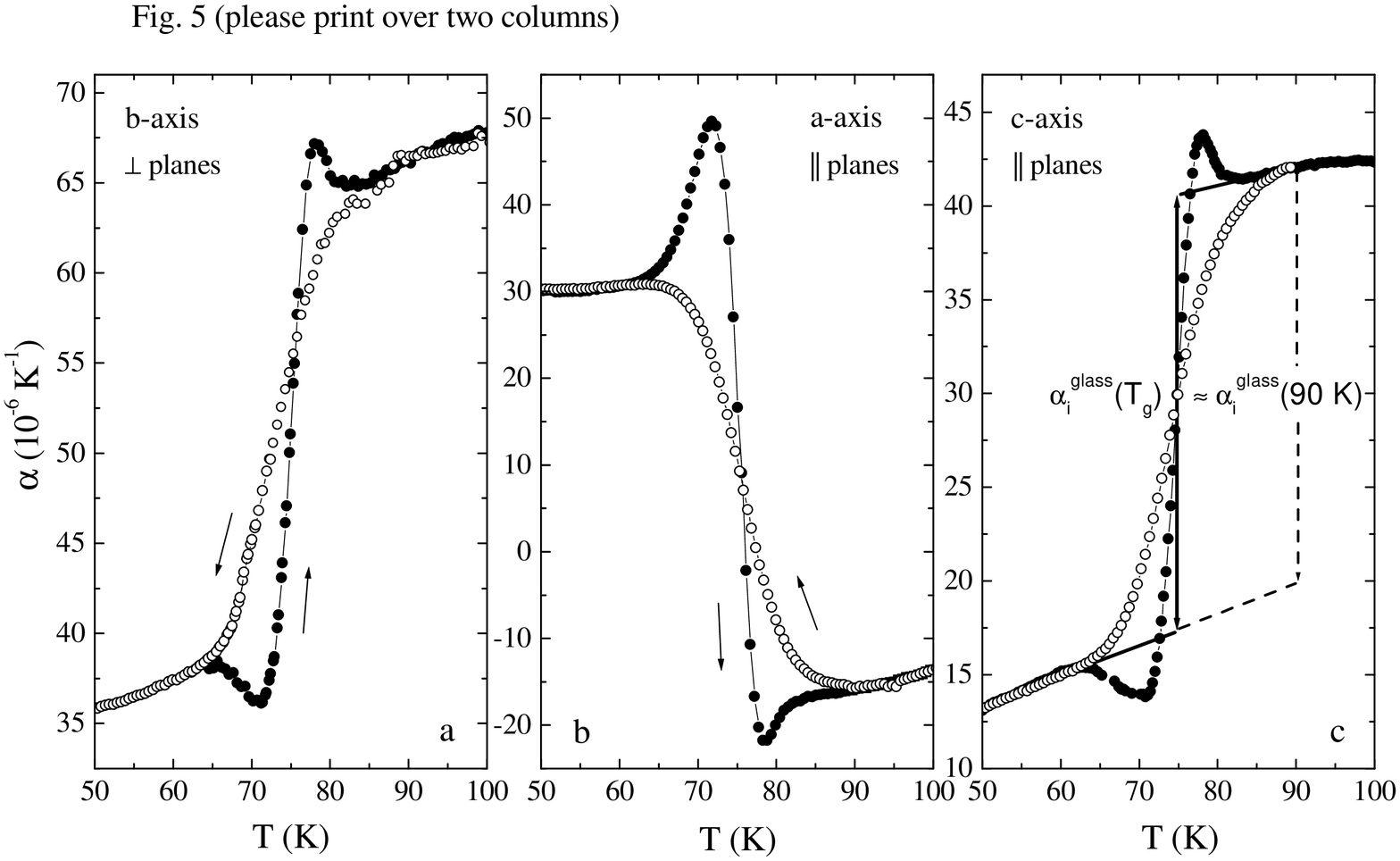,width=\textwidth}}
\vspace{0.5cm} \caption{Linear thermal expansion coefficients
$\protect\alpha_i(T)$ for the three principal axes of
$\protect\kappa$-(ET)$_2$Cu[N(CN)$_2$]Br in the vicinity of $T_g$.
Note the different vertical scales. Closed and open symbols denote
heating and cooling curves, respectively. In (c), the construction
to estimate the additional glassy contribution just above $T_g$ is
indicated (see text). The $\protect\alpha_b$ data (a) have been
taken on crystal \#\,2 while crystal \#\,3 was used for
$\protect\alpha_a$ (b) and $\protect\alpha_c$ (c).} \label{glass4}
\end{figure}

Fig.\ \ref{glass4} displays the linear thermal expansion
coefficient for the three principal axes, $\alpha _{i}(T)$, of
$\kappa $-H$_{8}$-Br. In accordance with the low symmetry of the
crystal structure, the lattice response at $T_{g}$ is strongly
anisotropic (note the different scales of the ordinates). The
additional glass contribution to the volume expansivity above
$T_{g}$, $\beta ^{{\rm glass}}(T)=\sum_{i}\alpha _{i}^{{\rm
glass}}(T)$, caused by the excitation of the motional
[(CH$_{2}$)$_{2}$] degrees of freedom is thermodynamically related
to the pressure dependence of the
%degree of ethylene order at a
%constant temperature. The latter is characterized by the molar
%entropy associated with this ordering process, $S_{\rm ethylene}$:
entropy associated with the ordering of the ethylene endgroups:
\begin{equation}
\left. \frac{\partial S_{{\rm ethy}}}{\partial p_{i}}\right|
_{T}=-V_{{\rm mol}}\cdot \alpha _{i}^{{\rm glass}}(T).
\label{eq:entropie}
\end{equation}
According to room-temperature X-ray diffraction studies, the
terminal ethylene groups for the $\kappa $-Br and $\kappa $-Cl
salts are disordered with the tendency towards the eclipsed
orientation \cite{Kini90}. The crystal structure taken at
127\thinspace K revealed an ordering in the eclipsed conformation
\cite{Geiser91} which was confirmed to be the case also at
20\thinspace K \cite{Geiser91a}. Since the X-ray diffraction
measurements give only an average structure, disorder at the
10\,\% level cannot be resolved.
%the disorder is not refined if there are less than about 10\,\% of
%the ethylene moieties in one configuration.
So, below 100\,K there may still be a considerable degree of
ethylene disorder. Since the background expansivity is unknown,
the temperature dependence of the additional glassy contribution
is difficult to determine. Yet, the discontinuities in the $\alpha
_{i}$'s can be used to estimate this contribution just above the
transition temperature where equilibrium can be established, i.e.,
at about 90\thinspace K: $\alpha _{i}^{{\rm glass}}(90\,{\rm
K})\approx \alpha _{i}^{{\rm glass}}(T_{g})=\alpha
_{i}(T\rightarrow T_{g}^{+})-\alpha _{i}(T\rightarrow T_{g}^{-})$.
This procedure, as exemplarily indicated in Fig.\ \ref{glass4}(c)
for the $c$-axis, yields $\alpha _{b}^{{\rm glass}}=+20\cdot
10^{-6}\,{\rm K}^{-1}$, $\alpha _{a}^{{\rm glass}}=-52\cdot
10^{-6}\,{\rm K}^{-1}$ and $\alpha _{c}^{{\rm glass}}=+22\cdot
10^{-6}\,{\rm K}^{-1}$ \cite{comment2}. Using equation
(\ref{eq:entropie})
and $V_{{\rm mol}}=499.37\,{\rm cm}^{3}$, we estimate $\partial S_{{\rm ethy}%
}/\partial p_{b}=-1$\thinspace J/(mol\ K\ kbar), $\partial S_{{\rm ethy}%
}/\partial p_{a}=+2.6$\thinspace J/(mol\ K\ kbar) and $\partial S_{{\rm ethy}%
}/\partial p_{c}=-1.1$\thinspace J/(mol\ K\ kbar). Consequently, pressure
along the $b$- and $c$-axes should increase the degree of [(CH$_{2}$)$_{2}$]
order, whereas for uniaxial stress along the $a$-axis the opposite effect is
expected. The uniaxial-pressure effects for stress along the in-plane $a$-
and $c$-axes are in accordance with the expectation:
%At least for the in-plane $a$- and $c$-axes these
%results may reflect the peculiarities of the crystal structure:
in the eclipsed conformation, the ethylene moieties of the donor molecules
making short C-H\thinspace $\cdots $\thinspace H and C-H\thinspace $\cdots $%
\thinspace anion contacts are aligned parallel to the $a$-axis \cite%
{Geiser91}. Therefore, stress along the orthogonal in-plane
$c$-axis should stabilize this conformation and thus should
increase the degree of order. On the other hand, the position of a
few still disordered ethylenes in the staggered conformation
should be stabilized for $a$-axis stress, as the alignment of the
staggered [(CH$_{2}$)$_{2}$] endgroups is parallel to $c$ (see,
e.g., Fig.\ 8 and Fig.\ 9 in \cite{Geiser91}). Combining the
uniaxial-pressure coefficients we find $\beta^{{\rm glass}}
=-10\cdot 10^{-6}\,{\rm K}^{-1}$ and $\partial S_{{\rm
ethy}}/\partial p_{hydr}=\sum_{i}\partial S_{{\rm ethy}}/\partial
p_{i}=+0.5$\thinspace J/(mol\ K\ kbar), i.e., a reduction of the
degree of ethylene order upon the application of hydrostatic
pressure. In a simple two-level model, the difference in the molar
entropy between the totally ordered and totally disordered
ethylene conformations amounts to $S_{{\rm ethy}}^{{\rm
max}}=Nk_{B}\ln 2=2R\ln 2=11.53$\thinspace J/(mol\ K) where
$N=2N_{A}$ denotes the number of relevant ethylene moieties per
mole. Given that only a few per cent of the ethylenes are still
disordered at $T_{g}$, i.e., that the major part of $S_{{\rm
ethy}}^{{\rm max}}$ has already been released at higher
temperatures, the above estimated value of $\partial S_{{\rm
ethy}}/\partial p_{hydr}$ underlines the extraordinary large
pressure/volume dependence of the ethylene-ordering effects in
$\kappa$-Br.

In order to estimate the activation energy $E_{{\rm a}}$ for the
$\kappa$-Cl salt, the cooling-rate dependence of $T_{g}$ was
measured in the same way as described above for the $\kappa $-Br
salt. Again, $T_{g}^{-1}$ follows well a linear $\ln |q_{c}|$
dependence in the Arrhenius plot yielding $E_{{\rm a}}=(2650\pm
300)$\thinspace K and $\nu_{0}=2 \times 10^{13 \pm 1.5}\,{\rm
Hz}$. The value for $E_{{\rm a}}$ agrees well with those reported
in the literature: $(2600\pm 100)$\thinspace K from $^{1}$H-NMR
\cite{Miyagawa95} and 2700\thinspace K from ac-calorimetry
\cite{Akutsu00}. Due to the particular shape of the $\kappa $-Cl
crystal used in this work, $\alpha $ could be measured only along
the interlayer direction, $\alpha _{b}$, and one in-plane
diagonal, $\alpha _{d}$. The values found for the discontinuities
at $T_{g}$ are $\Delta \alpha _{b}=+3.75\cdot 10^{-6}\,{\rm
K}^{-1}$ and $\Delta \alpha _{d}=-3.5\cdot 10^{-6}\,{\rm K}^{-1}$.
With $\alpha _{d}=\alpha _{a}\cos ^{2}\phi +\alpha _{c}\sin
^{2}\phi $ for the present orthorhombic crystal structure and
$\phi =45^{\circ }$, we get $\Delta \alpha _{a}+\Delta \alpha
_{c}=2\times \Delta \alpha _{d}$ yielding $\Delta \beta \approx
-3.25\cdot 10^{-6}\,{\rm K}^{-1}$ \cite{comment2}. Using $V_{{\rm
mol}}=496.66\,{\rm cm}^{3}$, we find $\partial S_{{\rm
ethy}}/\partial p_{hydr}=+0.16$\thinspace J/(mol\ K\ kbar), a
value which is significantly smaller than that found for $\kappa
$-Br. The correspondingly smaller pressure/volume effects on the
[(CH$_{2}$)$_{2}$] disorder may be due to the fact that the short
C-H\thinspace $\cdots $\thinspace H and C-H\thinspace $\cdots
$\thinspace anion contacts are stronger strained for the $\kappa
$-Cl salt which corresponds to a harder lattice \cite{Geiser91}.

\subsubsection{$\protect\kappa$-(ET)$_2$Cu(NCS)$_2$}
The shape of the glass-like transition for the monoclinic $\kappa $-Cu(NCS)$%
_{2}$ salt differs substantially from that of the orthorhombic
$\kappa $-Br and $\kappa $-Cl salts. Although these compounds are
similar in sharing polymeric, ribbon-like anion chains, the donor
molecules of the $\kappa $-Br and $\kappa $-Cl compounds lean
along the anion chain while they are perpendicular to the anion
chains for the $\kappa $-Cu(NCS)$_{2}$ salt. This results in a
different network of short C-H\thinspace $\cdots $\thinspace donor
and C-H\thinspace $\cdots $\thinspace anion contacts which
supports an ethylene-endgroup ordering in the staggered instead of
the eclipsed conformation \cite{Whangbo90}. Fig.\ \ref{glass1}(c)
clearly shows that in $\kappa $-Cu(NCS)$_{2}$, {\em two} closely
spaced glass-like transitions occur with small but distinct
anomalies around $T_{{\rm g}_{1}}=70$\,K and $T_{{\rm
g}_{2}}=53$\,K indicating that, in this case, a simple two-state
model is not adequate. Rather it appears that the transition
occurs in two steps characterized by %at least two channels for the
%ordering process exist, with
different activation energies $E_{\rm a}$. Due to the fact that
the cooling-curve anomalies can hardly be distinguished from the
unknown background expansivity, our data do not allow for a
reliable determination of $T_{g}(q_{c})$ and thus $E_{{\rm a}}$.
Although the freezing-in process seems to be more complicated as
it occurs in two steps, it is likely that also for $\kappa
$-Cu(NCS)$_{2}$ the ethylene endgroups of the donor molecules are
the relevant relaxation units. Concerning the effect on the volume
expansivity at $T_{g}$, we find that the uniaxial effects nearly
compensate each other so that the hydrostatic-pressure dependence
of the degree of ethylene ordering is rather small. This may be
due to the fact that the lattice of $\kappa $-Cu(NCS)$_{2} $ is
harder than that of $\kappa $-Br \cite{Whangbo90}. The rather weak
lattice response to the structural rearrangement process occurring
at $T_{g}$ indicates that the occupancy of the staggered
conformation is almost complete, i.e., the frozen-in degree of
disorder at low temperatures appears to be small for $\kappa
$-Cu(NCS)$_{2}$ compared to that for the $\kappa$-Br and $\kappa
$-Cl salts.

\subsection{Discussion}\label{glas2} %\subsubsection{$\kappa$-(ET)$_2$Cu[N(CN)$_2$]Br and
%$\kappa$-(ET)$_2$Cu[N(CN)$_2$]Cl}
The size of $T_{g}$ suggests that the energy difference $E_{{\rm
S}}$ between the eclipsed and staggered conformation is quite
small, i.e.\ of the order of about 100\,K. Both $E_{{\rm S}}$ as
well as the energy barrier $E_{{\rm a}}$ depend on the details of
the crystal structure, i.e., how the ethylene groups interact with
the anion and neighboring ET molecules, and also on how the charge
is distributed on the ET molecule because of the electrostatic
interaction between the positive protons and the electronegative
anions. Considering the slightly smaller unit-cell volume of
$V_{\rm uc} = 3223$\,\AA$^3$ (at 127\,K) for $\kappa$-Cl compared
to 3243\,\AA$^3$ for $\kappa$-Br \cite{Geiser91} as a slightly
higher chemical pressure for the former compound, the large
positive values of $\partial S_{{\rm ethy}}/\partial p_{hydr}$
estimated for this salt suggest that the degree of ethylene
disorder is higher for the insulating $\kappa $-Cl salt. A rough
estimate gives an entropy difference of $(0.2-0.3)$\,J/(mol\ K)
between the two salts which is significant compared to the
approximated estimated value of $S_{{\rm ethy}}(T_{g}) \approx
1$\,J/(mol\ K).

The transfer integrals $t_{{\rm eff}}$ between adjacent ET
molecules determining the electronic structure, depend strongly on
the intermolecular distance. Thus, changes in the lattice
parameters due to the above-identified glass-like transition may
cause anomalies in physical quantities which depend on the
transfer integrals as, e.g., transport properties. Indeed, a
pronounced kink in the resistivity at 75\thinspace K accompanied
by hysteresis between heating and cooling has been reported for the $\kappa $%
-H$_{8}$-Br salt \cite{Watanabe91}, consistent with our thermal expansion
results.\newline
Besides the kink structure in the resistivity, the H$_{8}$ and D$_{8}$ salts
of $\kappa $-Br show interesting time dependencies affecting both the
electronic properties at temperatures below $T^{\ddag }\approx (75-80)$%
\thinspace K and the actual ground-state properties. For superconducting $%
\kappa $-Br, Su {\em et al.} reported relaxation effects in $R(T)$
and a separation of the curves below 80\thinspace K as a function
of the cooling rate $q_{c}$\cite{Su98a,Su98b}. The way of cooling
through 80\thinspace K was found to influence the low-temperature
properties such that $T_{c}$ decreases on increasing $|q_{c}|$.
These phenomena have been ascribed to a structural transformation
leading to lattice disorder, whereas a glass-like ethylene
ordering was excluded \cite{Su98a,Su98b}. It was claimed that a
structural phase transition occurs in the anion system and that
fast cooling freezes in a high-temperature state where a
localization of charge carriers leads to localized magnetic
moments that suppresses $T_{c}$\cite{Su98c}. We note that the
quenching rates which were found to influence $T_{c}$ in those
studies ($|q_{c}|\geq 600$\,K/h) are much higher than the cooling
rates employed in our experiments.
%: up to $q_{c}\approx -300$\,K/h
%we observed neither a shift of $T_c$ nor a change of the shape of
%the phase-transition anomaly in $\alpha (T)$ at $T_{c}$.
A sequence of first-order phase transitions around 75\,K due to
ethylene-endgroup ordering was claimed based on resistance
measurements of structural relaxation kinetics \cite{Tanatar99}.
Magnetization measurements revealed that a growing amount of
disorder with increasing values of $|q_{c}|$ leads to larger
penetration depths and lower superconducting transition
temperatures \cite{Aburto98}. Besides these cooling-rate dependent
effects on $T_{c}$ for $\kappa $-H$_{8}$-Br, it has been reported
that rapid cooling through $T^{\ddag }\lesssim 80$\thinspace K
drives the superconducting ground state of the deuterated salt
$\kappa $-D$_{8}$-Br into an insulating antiferromagnetic state
\cite{Kawamoto97,Ito00}. This has been taken as a strong
indication for $\kappa $-D$_{8}$-Br being located right at the
border between a superconducting and a Mott-insulating
phase.\newline However, the above temperature $T^{\ddag }$, below
which time dependencies affecting the ground-state properties
become important, coincides with the actual glass-transition
temperature $T_{g}$. This suggests that the cooling-rate dependent
metal-insulator transition in $\kappa $-D$_{8}$-Br and the shift
of $T_{c}$ in $\kappa $-H$_{8}$-Br are related to the above
relaxation phenomena.

As the motional degrees of freedom of the [(CH$_{2}$)$_{2}$] units
become frozen-in below $T_{g}$ a {\em direct} interaction, i.e.\ a
scattering of the charge carriers off that motion is not expected.
%the number of ethylene moieties changing their
%conformational position per second, i.e.\ the relaxation rate, is
%given by the inverse of the relaxation time $\tau$, see equation
%(\ref{eq:tau}). According to this, $\tau^{-1}$ decreases
%exponentially with temperature and becomes practically zero at low
%temperatures.\\
Rather at $T_{g}$, the freezing-in process introduces, via the
C-H\thinspace $\cdots $\thinspace donor and C-H\thinspace $\cdots
$\thinspace anion contact interactions, a {\em random potential}
that may influence the effective transfer integrals $t_{{\rm eff}}$.
%is {\em randomly} altering the
%intermolecular contact distances with the C-H bonds and thus {\em
%indirectly} changing the electronic properties represented by
%$t_{\rm eff}$ as the C-H\,$\cdots$\,donor and C-H\,$\cdots$\,anion
%contact interactions are crucial for the packing arrangement
%\cite{Whangbo90}.
%Hence, $|q_c|$-dependent variations of the effective transfer
%integrals may influence not only the temperature dependence of the
%resistivity around $T_g$ but also the ground-state properties.
In fact, the comparison of two recent specific heat experiments on
$\kappa$-Br give strong indications for cooling-rate dependent
disorder: Nakazawa and Kanoda used the specific heat data of a
rapidly cooled deuterated crystal (which is insulating) to
determine the lattice specific heat for the $\kappa$-Br salt
\cite{Nakazawa97}. The so-derived lattice contribution was found
to differ substantially from the one observed in slowly-cooled
(superconducting) $\kappa$-H$_{8}$-Br \cite{Elsinger00}.\\
%indicating that the estimation of the phonon specific heat from a
%quenched-cooled deuterated sample is not approriate.
It is obvious to assume that the random potential induced by large
cooling rates is unfavourable for superconductivity in the present
materials \cite{Geiser91}. One can think of different mechanisms
as to how the disorder affects the superconducting state: (i)
Extended structural defects may act as non-magnetic scattering
centers but can also have an effect on the density of states
$N(E_{F})$ due to the deformation of the lattice. (ii) Disorder
can induce local magnetization and thus magnetic scattering (pair
breaking). In a recent theoretical study, it was pointed out that
the effect of disorder on the interplay between magnetism and
superconductivity can be crucial for strongly correlated electron
systems \cite{Kohno99}. (iii) Lattice disorder can lead to
Anderson localization and metal-insulator transition, in
particular in low-dimensional systems.\newline To clarify the
relative role of the above effects, further material-specific
experimental and theoretical studies are necessary. From our
experiments we only can state that the phenomenology of the
glass-like transition in the ethylene-endgroup degrees of freedom
leading to lattice disorder provides a natural explanation for the
time dependencies and cooling-rate dependent effects reported in
literature. To give some numbers, for the deuterated $\kappa
$-D$_{8}$-Br salt, an increase of the cooling rate to 100\,K/min
has been reported to cause a $13\,\%$ depression of $T_{c}$ (from
11.5\thinspace K to 10\thinspace K) and a substantial reduction of
the superconducting volume fraction compared to slow cooling
(0.2\thinspace K/min) \cite{Kawamoto97}. According to our results
(assuming $E_{{\rm a}}$ to be identical to that for $\kappa
$-H$_{8}$-Br) such an increase in $|q_c|$ is expected to cause an
$18\,\%$ shift of $T_{g}$ to higher temperatures (from
79\thinspace K to 93\thinspace K). This would correspond to a
considerably higher degree of frozen-in disorder in the rapidly
cooled sample which may qualitatively explain the depression of
$T_c$.

To disclose the role of ethylene disorder for the superconducting
properties, comparative pressure experiments with hydrostatic
pressure applied at temperatures well above and below $T_g$ would
be helpful: due to the large pressure/volume dependence of
$S_{{\rm ethy}}$, the degree of disorder is considerably increased
for the former experiment compared to the latter, even for a small
hydrostatic pressure.
%If the pressure values (not too large, so
%that the sample remains superconducting) and the thermal histories
%would be the same in both cases, the current state of the sample
%would differ with respect to the degree of disorder which could be
%controlled by measuring, e.g., the residual resistivity ratio.
Subsequent measurements of the Meissner and shielding volume and
the resistivity could provide valuable information about the
implication of lattice disorder on the superconducting state of
$\kappa$-Br. According to this, the latter compound may be
considered as a reference system, where the influence of
structural disorder on the superconducting properties can be
studied in a controlled way on the {\em same} sample.

\section{The antiferromagnetic transition in insulating $\protect%
\kappa$-$(\mbox{ET})_2\mbox{C\lowercase{u}}[\mbox{N}(\mbox{CN})_2]
\mbox{C\lowercase{l}}$} \label{AFI} Although being isostructural
to the superconducting $\kappa $-Br salt, the small modification
of the anion composition of $\kappa $-Cl leads to subtle changes
of the donor arrangement and to an insulating low-temperature
state. Early magnetic measurements revealed the onset of a shallow
decrease in the magnetization upon cooling to below 45\thinspace
K, which was regarded as a signature of antiferromagnetic ordering
\cite{Welp92}. In addition, these studies revealed indications for
a weakly ferromagnetic state with a small saturation moment of
$8\cdot 10^{-4}\,\mu _{B}/{\rm dimer}$ below 22\thinspace K.
Subsequently, the spin structure has been studied by $^{1}$H-NMR
measurements \cite{Miyagawa95}, yielding a commensurate
antiferromagnetic order at $T_{N}=27$\thinspace K with a moment of
$(0.4-1.0)\,\mu _{B}/{\rm dimer}$. From these measurements, the
authors inferred that the easy axis of the ordered moments is
perpendicular to the layers and that a small canting of the spins
below 23\thinspace K gives rise to a ferromagnetic moment parallel
to the layers.\\ Fig.\ \ref{motivation}(a) clearly demonstrates
that $\alpha _{\perp}(T)$ of $\kappa $-Cl exhibits another
distinct feature besides the glass-like transition: a negative
second-order phase-transition anomaly slightly below 28\,K. As the
anomaly in $\alpha _{\perp}(T)$ occurs at the same temperature
below which $^{1}$H-NMR measurements revealed the onset of
magnetic order, we regard the $\alpha _{\perp}(T)$ jump as the
bulk signature of the antiferromagnetic transition.

Three different proposals have been made for the origin of the
magnetic moments and the character of the magnetic state in
$\kappa $-Cl; (i) lattice disorder due to conformational
ethylene-endgroup disarrangements \cite{Posselt94} - here the
localization of electron states with incomplete compensation of
their spins is believed to cause an inhomogeneous, frozen-in
magnetic state at low temperatures; (ii) nesting properties of the
Fermi surface giving rise to itinerant, spin-density-wave (SDW)
magnetism \cite{Wzietek96,Tanatar97}; (iii) a Mott-Hubbard type
metal-insulator transition leading to a magnetic state
characterized by localized spins \cite{Miyagawa95}.\\ As our
measurements provide clear thermodynamic evidence for a phase
transition at $T_{N}$, the first proposal can be discarded. To
check for the proposals (ii) and (iii) it is be helpful to inspect
the anisotropies in the $\alpha (T)$ response at $T_{N}$ that
allows for a determination of the uniaxial-pressure effects on
$T_{N}$. Within our experimental resolution, there is no anomaly
visible at $T_{N}$ for both in-plane thermal expansion
coefficients, i.e.\ $\Delta \alpha _{\parallel } \simeq 0$ (not
shown) \cite{comment4}. According to the Ehrenfest relation, eq.\
(\ref{eq:ehrenfest}), this corresponds to a vanishingly small
in-plane pressure effect on $T_{N}$. Taken together, the negative
pressure coefficient of $T_{N}$ for uniaxial stress perpendicular
to the planes and the negligible in-plane-pressure effect imply a
negative pressure effect on $T_{N}$ under hydrostatic-pressure
conditions, in agreement with the experimental observations.\\ In
contrast to the above dilatometric studies, specific heat
measurements have failed so far to detect a phase-transition
anomaly \cite{Nakazawa96}. Using the jump height for the volume
expansivity $\Delta \beta \simeq \Delta \alpha _{\perp } = -2
\cdot 10^{-6}\,{\rm K}^{-1}$ and available literature data on the
response of\ $T_{N}$ to hydrostatic pressure varying between
$-150$\,K/kbar \cite{Schirber91} and $-25$\thinspace K/kbar
\cite{Lefebvre00}, the Ehrenfest relation allows to estimate the
expected discontinuity in the specific heat at $T_{N}$. We find
$\Delta C_{m}\approx (20-100)\,{\rm mJ/(mol\ K)}$, which is much
below the experimental resolution of the specific heat
measurements reported in \cite{Nakazawa96}.

The above findings of a highly anisotropic lattice response at
$T_N$ and, related to this, strongly directional-dependent
uniaxial-pressure coefficients, provide a crucial test for models
attempting to describe the nature of the antiferromagnetism in
$\kappa$-Cl. The lack of a visible $C(T)$ anomaly has been
ascribed to the 2D character of the spin correlations, resulting
in a short-range 2D ordering of the spins well above the 3D
transition temperature. The magnetic exchange-coupling constant
for nearest-neighbor interactions was estimated to $J^{\parallel
}\sim 460$\thinspace K \cite{Nakazawa96}. Accordingly, most of the
entropy of $R\ln 2$ is released at temperatures far above $T_{N}$
This situation is strikingly similar to that in La$_{2}$CuO$_{4}$
where the expected jump height in specific heat at the 3D magnetic
ordering, $\Delta
C_{m}(T_{N})$, was %in the order of $[S_{m}(T_{\rm
%N})/S_{m}(\infty)]\Delta C_^{\rm MF}{m}$ where $S_{m}(T)$ denotes
%the magnetic entropy ($S_{m}(\infty) = R \ln
%2$) and $\Delta C_^{\rm MF}{m}$ the mean field value (1.5\,R)
%which was estimated to be
below the experimental resolution due to the strong 2D antiferromagnetic
correlations with $J^{\perp }/J^{\parallel }\approx 10^{-5}$ \cite{Sun91}, $%
J^{\perp }$ being the interplane coupling constant. Apart from the small
jump in the specific heat which would be compatible with a 3D
antiferromagnetic ordering among localized spins at $T_{N}$, i.e.\ proposal
(iii), this model is difficult to reconcile with the distinct anisotropy in $%
\partial T_{N}/\partial p_{i}$ deduced from our measurements. On the one
hand, for such a scenario, one would expect that in-plane pressure
affecting predominantly $J^{\parallel }$ should also influence the
3D ordering temperature. This contrasts with our finding of a
vanishingly small $\partial T_{N}/\partial p_{\parallel}$. On the
other hand, from the negative pressure coefficient of $T_{N}$ for
uniaxial stress perpendicular to the planes we infer $\partial
J^{\perp }/\partial p_{\perp }<0$. However, a decrease of the
interlayer coupling constant upon reducing the interlayer distance
cannot be understood regarding only nearest-neighbor magnetic
couplings in a 3D spatially anisotropic Heisenberg model. The
latter has been successfully applied to explain the experimental
data of La$_2$CuO$_4$ \cite{Siurakshina00}. However, for the
complex crystal structures of the title compounds there are
additional, frustrating magnetic couplings \cite{McKenzie98}.
Possibly, these are relevant not only for the in-plane but also
for the out-of-plane directions. A suppression of the magnetic
order upon stress perpendicular to the planes could be understood
if those frustrating interactions were to increase more strongly
than the nearest-neighbor couplings. For in-plane pressure, both
interactions would have to just cancel out in our case. An
explanation in terms of frustrating magnetic couplings could
account also for the low value of the transition temperature
$T_N$.\\ With respect to a SDW scenario, i.e.\ proposal (ii), the
following aspects are of relevance: as it is likely that the
nesting vector lies normal to the open sheets of the Fermi surface
with the largest component along the $c$-axis in the conducting
plane, in-plane stress is expected to affect the FS topology and
thus the nesting properties. In fact, recent uniaxial-strain
studies of the SDW transition in $\alpha
$-(ET)$_{2}$KHg(SCN)$_{4}$ revealed that uniaxial pressure along
both in-plane axes alter the nesting properties and, thereby,
cause a strong suppression of $T_{SDW}$ \cite{Maesato01}. Thus,
the absence of any in-plane pressure effect on the
antiferromagnetic phase transition in $\kappa $-Cl appears to be
in conflict with a simple spin-density-wave scenario.
Nevertheless, the fact that uniaxial pressure along the
out-of-plane direction suppresses $T_{N}$ (and subsequently
induces superconductivity) would be compatible with a SDW
scenario. Uniaxial stress perpendicular to the planes is expected
to increase the warping of the cylindrical FS which in turn
reduces the nesting properties, thereby allowing superconductivity
to form. In fact, such a behavior is observed for
$\alpha$-(ET)$_{2}$KHg(SCN)$_{4}$ \cite{Campos95,Maesato01}.\\ %For a
%SDW transition, the mean-field description predicts for
%discontinuity in the specific heat, that $\Delta C_{m}/\gamma
%T_{{\rm SDW}}=1.43$. Using the value of $\gamma =25\,{\rm mJ/(mol\
%K^{2})}$ for the isostructural, metallic $\kappa$-Br system as an
%upper limit \cite{Elsinger00}, one gets $\Delta C_{m}\approx
%1\,{\rm J/(mol\ K)}$ which would also be consistent with the
%specific heat measurements reported in \cite{Nakazawa96}.\\
Our results indicate that none of the above proposals is suited to
describe the magnetic state in $\kappa$-Cl satisfactorily.
However, certain elements of both models, i.e.\ of a magnetism of
localized moments like it is found for the high-$T_c$ cuprates and
of an itinerant, nesting-driven magnetism like in the Bechgaard
salts seem to apply for $\kappa$-Cl. This suggests a more
complicated magnetic behavior than supposed up to now. In section
\ref{45K} we will discuss a spin-density-wave scenario as for the
origin of the phase-transition-like anomalies at $T^{\ast}$ in the
superconducting salts. In this context we will speculate that the
same magnetic interactions cause an antiferromagnetic order among
localized spins in insulating $\kappa$-Cl.

\section{Instabilities in superconducting $\protect\kappa$-$(
\mbox{ET})_2\mbox{C\lowercase{u}}[\mbox{N}(\mbox{CN})_2]\mbox{B\lowercase{r}}$
and
$\protect\kappa$-$(\mbox{ET})_2\mbox{C\lowercase{u}}(\mbox{NCS})_2$}
\label{normalstate}
\subsection{Uniaxial-pressure coefficients of $T_c$}
\label{uniaxial} Fig.\ \ref{intermedia} shows the linear thermal
expansion coefficients $\alpha _{i}(T)$ for $T\leq 60$\thinspace K
of (a) $\kappa $-Br and (b) $\kappa $-Cu(NCS)$_{2}$. For both
compounds the lattice response at $T_{c}$ is strongly anisotropic.
In a comparative thermal expansion study aiming at a determination
of the uniaxial-pressure coefficients of $T_{c}$ we found that the
latter system exhibits a strikingly similar anisotropy of
$\partial T_{c}/\partial p_{i}$ as the compound $\beta ^{\prime
\prime}$-(ET)$_{2}$SF$_{5}$CH$_{2}$CF$_{2}$SO$_{3}$, namely either
a positive or vanishingly small pressure coefficient for uniaxial
pressure along both in-plane axes, but a huge negative interplane
coefficient $\partial T_{c}/\partial p_{\perp }$ \cite{Mueller00}.
Pronounced negative cross-plane pressure effects $\partial
T_{c}/\partial p_{\perp }<0$ were found also for the $\kappa
$-(ET)$_{2}$X salts with X=I$_{3}$ and Cu[N(CN)$_{2}$]Br
\cite{Mueller01}. The results for the latter salt are in conflict
with those reported by Kund {\em et al.} \cite{Kund94a}, where a
vanishingly small cross-plane pressure coefficient of $(0\pm
0.2)\,{\rm K/kbar}$ was claimed. Here we present high-resolution
measurements of the linear expansivities for the three principal
axes for the $\kappa $-Br salt which clearly reveal negative
discontinuities at $T_{c}$ along all three axes: $\Delta \alpha
_{b}=-(2.1\pm 0.3)\cdot 10^{-6}\,{\rm K}^{-1}$ (out-of-plane) and
for the in-plane $a$- and $c$-axes $\Delta \alpha _{a}=-(1.93\pm
0.3)\cdot 10^{-6}\,{\rm K}^{-1}$ and $\Delta \alpha _{c}=-(0.2\pm
0.08)\cdot 10^{-6}\,{\rm K}^{-1}$, respectively. The values for
the in-plane discontinuities agree well with those reported in a
previous thermal expansion study by Kund {\em et al}. In order to
rule out sample dependencies, we studied a second crystal from a
different batch and found almost identical behavior
\cite{Mueller01}. Using the Ehrenfest relation, eq.\
(\ref{eq:ehrenfest}), and the jump height $\Delta C$ reported in
literature \cite{Elsinger00a} we find for the uniaxial-pressure
coefficients of $T_{c}$ for the $\kappa $-Br system: $\partial
T_{{\rm c}}/\partial p_{b}=-(1.26\pm 0.25)\,{\rm K/kbar}$
(out-of-plane) and for the in-plane coefficients $\partial
T_{c}/\partial p_{a}=-(1.16\pm 0.2)\,{\rm K/kbar}$ and $\partial
T_{{\rm c}}/\partial p_{c}=-(0.12\pm 0.05)\,{\rm K/kbar}$. For the
hydrostatic-pressure coefficient we get $\partial T_{c}/\partial
p_{hydr}=\sum_{i}(\partial T_{c}/\partial p_{i})=-(2.58\pm
0.5)\,{\rm K/kbar} $ \cite{comment5}, in excellent agreement with
the values found in hydrostatic-pressure experiments, i.e.\
$-(2.4-2.8)\,{\rm K/kbar}$ \cite{Schirber90,Sushko91}. We stress
that these uniaxial-pressure results for the various (ET)$_2$X
superconductors {\em do not} yield a uniform behavior as for the
intralayer pressure effects on $T_c$. In particular, the results
on $\kappa$-Cu(NCS)$_2$ show that in-plane pressure can even cause
an increase of $T_c$ \cite{Mueller00}. Note that this is in
contrast to what is assumed in a 2D electronic model \cite{Kino96}
and demonstrates that an attempt to model these systems by solely
considering in-plane electronic properties is not appropriate.\\
However, a large negative {\em inter}layer pressure coefficient of
$T_{c}$ that predominates the hydrostatic pressure effect is
common to all $\kappa $-(ET)$_{2}$X salts investigated so far. As
we discussed in \cite{Mueller00,Mueller01} the sensitivity of
$T_{c}$ to changes of the cross-plane lattice parameter can arise
from pressure induced changes of both the interlayer interaction,
i.e., the strength of the 3D coupling and/or the vibrational
properties of the lattice.

\subsection{Phase-transition anomalies above $T_c$}
\label{45K}
\subsubsection{Phenomena and analysis}
\begin{figure}[tbp]
\centerline{\psfig{figure=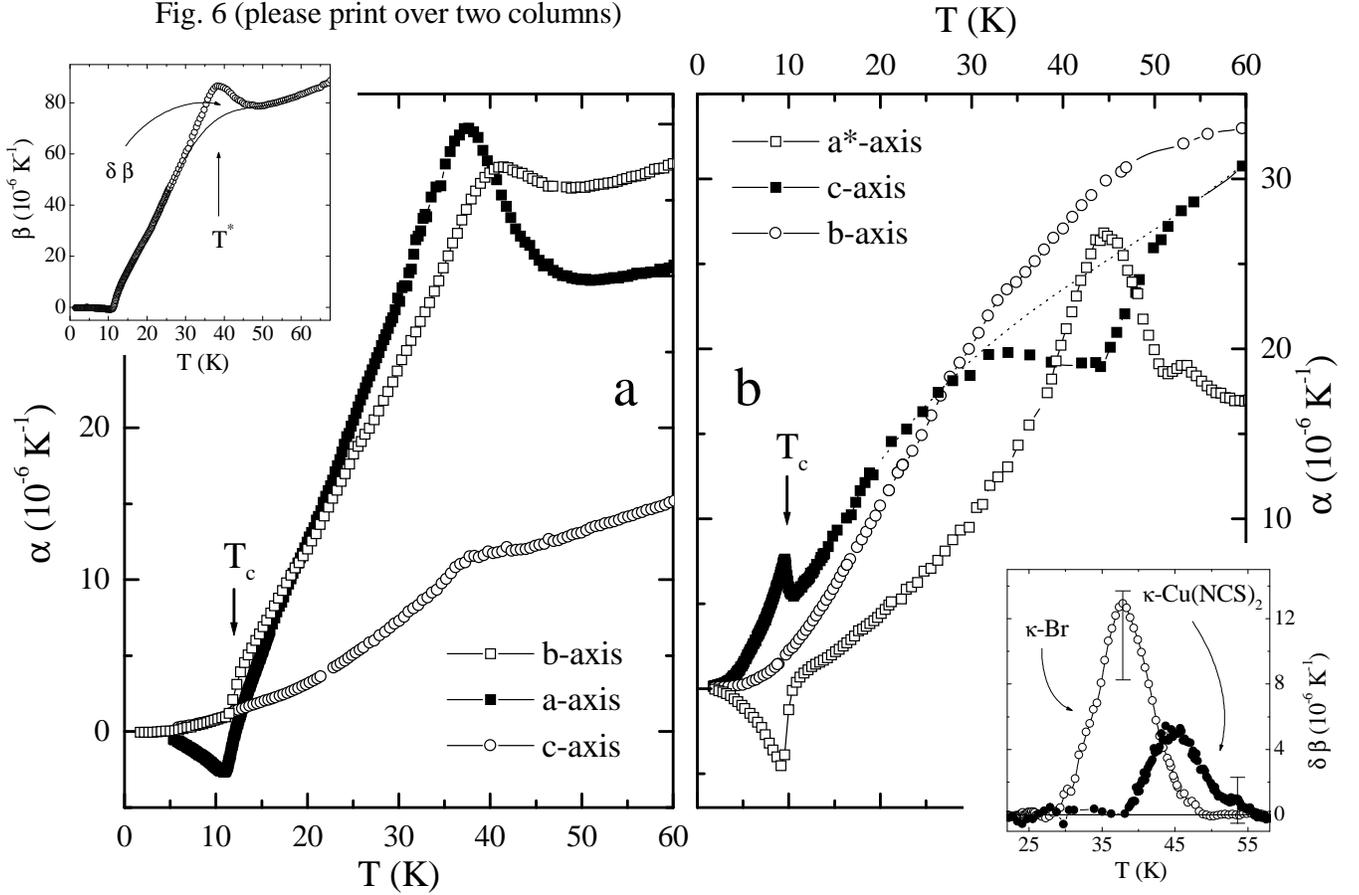,width=\textwidth}}
\vspace{0.5cm} \caption{Linear thermal expansion coefficients
$\protect\alpha_i(T)$ for the three principal axes of (a)
$\protect\kappa$-(H$_8$-ET)$_2$Cu[N(CN)$_2$]Br and (b)
$\protect\kappa$-(D$_8$-ET)$_2$Cu(NCS)$_2$. Same symbols indicate
related symmetry axes for both salts: the interlayer direction, i.e.\ the $b$%
-axis for the former and the $a^{\ast}$-axis for the latter salt, the
in-plane axis along which the polymeric anion chains run, i.e.\ the $a$- [$%
\protect\kappa$-Br] and $c$-axis [$\protect\kappa$-Cu(NCS)$_2$] and the
second in-plane axis perpendicular to the anion chains, i.e.\ $c$- [$\protect%
\kappa$-Br] and $b$-axis [$\protect\kappa$-Cu(NCS)$_2$]. The inset
in the left panel shows the volume thermal expansion coefficient
$\protect\beta (T)$ of $\protect\kappa$-Br. The solid line
indicates the interpolated lattice background. The inset in the
right panel compares the anomalous additional
contributions to the volume expansivity, $\protect\delta \protect\beta$, of $%
\protect\kappa$-Br and $\protect\kappa$-Cu(NCS)$_2$ estimated as explained
in the text. For $\protect\kappa$-Br (a) the $\protect\alpha_b$ data have
been taken on crystal \#\,2 while crystal \#\,3 was used for $\protect\alpha%
_a$ and $\protect\alpha_c$.}
\label{intermedia}
\end{figure}

Besides the discontinuities associated with the superconducting
instability in temperature region (A), the linear thermal
expansion coefficients of both compounds show unusual expansivity
behaviour in the range $30-50$\thinspace K (B). For $\kappa
$-H$_{8}$-Br, cf.\ Fig.\ \ref{intermedia}(a), we find anomalous
expansivity contributions $\delta \alpha _{i}$, i.e.\ a pronounced
local maximum and a shoulder around $37.5$\thinspace K for the
in-plane $a$- and $c$-axis, respectively, and a maximum at
41.5\thinspace K for the out-of-plane $b$-axis. The slight
differences in the characteristic temperatures of these anomalies
are most likely an artefact due to the unknown background
expansivities, which makes a separation of $\delta \alpha _{i}$
difficult. As will be shown below, the corresponding contribution
to the volume expansivity, $\delta \beta $, can be estimated in a
more reliable way allowing for a determination of the
characteristic temperature $T^{\ast}$. We note that for $\kappa
$-Br these anomalous contributions at $T^{\ast }$, $\delta \alpha
_{i}(T^{\ast})$, are positive along all three axes whereas the
discontinuities at $T_{c}$, $\Delta \alpha _{i}(T_{c})$, are all
negative. Furthermore, we find the smallest effect for $\alpha
_{c}$ where also the lattice response at $T_{c}$ is smallest. The
inset of Fig.\ \ref{intermedia}(a) shows the volume thermal
expansion coefficient $\beta (T)=\sum_{i}\alpha _{i}(T)$ of
$\kappa $-H$_{8}$-Br together with a smooth interpolation from
temperatures below and above the range of anomalous expansivity
behavior. This procedure allows for a rough estimate of the
anomalous contribution $\delta \beta (T)$.\\ At slightly higher
temperatures, anomalous expansivity behavior is found also for the
$\kappa $-Cu(NCS)$_{2}$ salt, see Fig.\ \ref{intermedia}(b). Here,
the background expansivity along the out-of-plane direction
$a^{\ast }$ is rather difficult to estimate due to both the
nearness of the glass-like anomaly around 53\thinspace K and the
unusual overall temperature dependence of the $a^{\ast }$ lattice
parameter. Yet, a determination of the sign and the approximate
size of the anomalies for each axis is still possible: there is
almost no anomalous contribution visible for $\alpha _{b}$ but a
distinct negative anomaly for $\alpha _{c}$, cf.\ the dotted line
representing the approximate background contribution. The largest
effect, i.e.\ a sharp maximum at $T^{\ast }=45$\thinspace K is
found for $\alpha _{a^{\ast }}$. Again, as for $\kappa
$-H$_{8}$-Br, the signs of these anomalies are strictly
anticorrelated with those of the discontinuities at $T_{c}$: there
is almost no anomaly in $\alpha _{b}$ both around $T^{\ast }$ and
at $T_{c}$; a positive jump in $\alpha _{c}$ at $T_{c}$ is
accompanied by a negative contribution at $T^{\ast }$ and the
large negative $\Delta \alpha _{a^{\ast }}$ at $T_{c}$ complies
with the large positive peak structure at $T^{\ast }$. These
results imply that for both compounds, the uniaxial-pressure
coefficients of the anomaly at $T^{\ast }$ and those at $T_{c}$
are strictly
anticorrelated in their signs but correlated in their magnitudes: for $%
\kappa $-Cu(NCS)$_{2}$, for example, uniaxial pressure applied
perpendicular to the planes causes a substantial shift of the
45\thinspace K anomaly to higher temperatures while at the same
time it strongly reduces $T_{c}$. The inset of Fig.\
\ref{intermedia}(b) compares the corresponding contributions to
the volume expansion coefficient for both salts. We find sharp
peaks in $\delta \beta $ at $T^{\ast }=38$\thinspace K and
45\thinspace K for $\kappa $-H$_{8}$-Br and $\kappa
$-Cu(NCS)$_{2}$, respectively, reminiscent of somewhat broadened
second-order phase transitions. We note that the overall shape of
the anomaly, i.e.\ its width and asymmetry, but not the peak
itself depends somewhat on the interpolation procedure employed to
determine the background expansivity. The error bars include the
uncertainties in the interpolated lattice background and
sample-to-sample variations.

\subsubsection{Discussion}
The above described anomalies in $\alpha (T)$ at intermediate
temperatures are particularly interesting because magnetic,
transport and elastic properties exhibit anomalous behavior in the
same temperature range. As mentioned in the introduction, the
presence of strong antiferromagnetic spin fluctuations preceding
the superconducting transition has been inferred from measurements
of the $^{13}$C-NMR \cite{Kawamoto95}, yielding a pronounced
maximum in $(T_1 T)^{-1}$ at around 50\,K for $\kappa$-Cu(NCS)$_2$
and at somewhat lower temperatures for the $\kappa$-Br salt. The
rapid decrease of $(T_1 T)^{-1}$ below 50 K has been ascribed to
the opening of a pseudogap in the spin-excitation spectrum
\cite{Kanoda97,Nakazawa00} - an interpretation that has been
proposed first by Kataev {\em et al.} based on their analysis of
ESR data \cite{Kataev92}. At about the same temperature, anomalous
behavior has been reported also from various other quantities, as,
e.g., the decrease of the spin susceptibility studied by
Knight-shift and dc-magnetization  measurements
\cite{Mayaffre94,Kawamoto95a}, indicative of a depression of the
density of states at the Fermi level and the resistivity, showing
a pronounced peak in its temperature derivative dR/dT
\cite{Sushko91,Gärtner88,Murata90}. The coincidence of these
features in the electronic and magnetic properties with the
$\alpha (T)$ anomalies described above suggest a common origin of
the various phenomena. This is corroborated also by the response
of these features to hydrostatic pressure. From the positive sign
of the anomalous expansivity contribution $\delta \beta (T)$, cf.\
inset of Fig.\ \ref{intermedia}(b), a shift of $T^{\ast}$ to
higher temperatures is expected under hydrostatic pressure. In
fact, a positive hydrostatic-pressure coefficient has been found
in the magnetic and transport measurements
\cite{Mayaffre94,Murata90,Sushko91}. A similar, positive pressure
effect has been observed also for the anomalies in the elastic
constants. At ambient pressure, a pronounced softening of
ultrasonic modes has been detected at 38\,K and 46\,K for
$\kappa$-Br and $\kappa$-Cu(NCS)$_2$, respectively
\cite{Frikach00,Simizu00}. These sound velocity anomalies have
been assigned to a magnetic origin due to their phenomenological
relation to the NMR results \cite{Frikach00}. However, an
alternative interpretation in terms of a structural phase
transition could not be excluded \cite{Simizu00}. Owing to the
large response of the anomaly in $\alpha (T)$, thermal expansion
measurements provide a most sensitive tool to investigate these
effects in detail. In particular, a comparative study covering
various related (ET)$_2$X compounds is helpful to find out the
relevant structural and/or electronic parameters involved.
Interestingly enough, corresponding features are absent in
$\beta''$-(ET)$_2$SF$_5$CH$_2$CF$_2$SO$_3$ ($T_c=5$\,K; large
discrete anions) \cite{Mueller00}, $\kappa$-(ET)$_2$I$_3$ ($T_c =
3.5$\,K; linear anions) as well as $\alpha$-(ET)$_2$KHg(SCN)$_4$
(non-superconducting; thick, polymeric anions) \cite{unpublished},
showing smooth, Debye-like temperature dependencies of $\alpha
(T)$ along all crystal axes for temperatures up to 200\,K.
Accordingly, purely {\em intra}molecular changes of the ET
molecule can be ruled out to account for the $T^\ast$ anomalies in
the $\kappa$-Br and $\kappa$-Cu(NCS)$_2$ salts. Obviously, the
peculiarities of the packing arrangement and the coupling to the
charge compensating anions, both of which result in a very similar
electronic structure for the latter two salts (but a slightly
different one for the various others), are crucially important for
the anomalies at $T^\ast$. An intimate relation to the electronic
structure is in line with the absence of corresponding features in
both the non-metallic $\kappa$-Cl as well as in the deuterated
$\kappa$-D$_8$-Br, cf.\ Fig.\ \ref{motivation}, and is also
corroborated by our supplementary investigations on the isotope
effect: For the $\kappa$-Cu(NCS)$_2$ salt we compared
$\alpha_{\perp}(T)$ of the pure system to that of crystals with
various isotope substitutions. This includes not only the
deuterated D$_8$ counterpart but also isotope substitutions at the
anion as well as the cation sites. For the latter, a
$\kappa$-(D$_8$$^{13}$C$_4$$^{34}$S$_8$-ET)$_2$Cu(NCS)$_2$ crystal
yielding a mass change of 28\,amu per ET molecule has been
investigated. For all isotopically labelled crystals the maximum
in $\alpha_{\perp}(T)$ could be reproduced in detail without any
significant shift in $T^\ast$. From the absence of a measurable
isotope shift, cooperative structural changes in the
ethylene-endgroups as well as in the anion or {\em inter}molecular
rearrangements in the cation system can be ruled out. Hence, we
propose that the anomaly at $T^\ast$ is caused by the electronic
structure, i.e.\ the Fermi-surface topology, which, according to
band structure calculations, should be similar for the $\kappa$-Br
and $\kappa$-Cu(NCS)$_2$ salts \cite{Whangbo90}. The presence of
strong antiferromagnetic interactions as seen in the NMR
experiments then hint at a spin-density-wave (SDW) rather than a
pure charge-density-wave (CDW) formation. This scenario implies
that the SDW and superconductivity form on disjunct parts of the
Fermi surface (FS). The SDW state is likely to involve the small,
quasi-onedimensional portions while leaving the major
twodimensional parts of the FS unaffected. The latter is subject
to the superconducting instability in the $\kappa$-Br and
$\kappa$-Cu(NCS)$_2$ salts. We suppose that a relative shift of
one of these FS portions in favour of the other can be induced by
the application of uniaxial pressure: For instance, uniaxial
pressure along certain crystallographic directions may destroy the
nesting properties of the quasi-onedimensional parts, thereby
destabilising the SDW state and causing a negative pressure
coefficient of $T^\ast$. As a consequence, the remaining
quasi-twodimensional parts of the FS will increase. Since this
will reinforce superconductivity, an increase of $T_c$ is
expected. On the other hand, those uniaxial-pressure conditions
that improve the nesting properties and, by this, cause an
increase in $T^\ast$ will result in a reduced remaining FS and,
thus, in a drop of $T_c$. Such an anticorrelation of the
uniaxial-pressure coefficients of $T^\ast$ and $T_c$ is exactly
what we have found in our experiment.\\
%According to these results, it is the uniaxial pressure
%perpendicular to the conducting planes that improves the nesting
%properties of the FS most strongly, and, by this destabilizes
%superconductivity.
The presence of two "subsets" of $\pi$-electrons in \boldmath
$k$\unboldmath-space corresponding to the quasi-1D and -2D parts
of the FS would be compatible with the NMR results: while the
quasi-2D parts cause an ordinary Korringa-type relaxation
behavior, the good nesting properties of the quasi-1D portions
give rise to a $(T_1 T)^{-1}$ component that grows with decreasing
$T$. Upon cooling through $T^{\ast}$, a real gap on that part of
the FS opens and causes a rapid drop in the corresponding $(T_1
T)^{-1}$ contribution. A loss of the minor part of the FS (say
about 10\,\%) is also capable to account for the change in the
temperature dependence of the resistivity from a larger slope ($T
> T^\ast$) to a smaller slope ($T < T^\ast$), i.e.\ a relative
increase in $\rho(T)$ at $T < T^\ast$. Such a scenario implies
that cooling through $T^\ast$ is accompanied not only by a
reduction of the spin susceptibility, as has been seen in several
of the above-mentioned experiments, but also by the development of
a (weak) magnetic anisotropy. To check for this possibility, we
have initiated a careful study of the spin susceptibility on the
various $\kappa$-(ET)$_2$X salts which, in fact, confirm our
expectation \cite{Sasaki01}.

It remains to be seen, whether the disappearance of a minor part
of the Fermi surface would be compatible with the results of the
Fermi-surface studies on these materials. If we speculate a little
further, it is tempting to associate the antiferromagnetism in the
$\kappa$-Cl salt also with a SDW state. As a consequence, we
propose that in the latter salt, the large gap seen in various
quantities such as the optical conductivity \cite{Kornelsen92},
involves only the charge degrees of freedom but leaves the spins
unaffected. Upon cooling down to $T_N = 27.8$\,K in such a
commensurate charge-ordered state, the same antiferromagnetic
interactions between nearest neighbour spins that cause the SDW
state in the $\kappa$-Br and $\kappa$-Cu(NCS)$_2$ salts then lead
to a commensurate antiferromagnetic order in $\kappa$-Cl.

\section{Conclusion}
\label{conclusion} 
\begin{figure}[tbp]
\centerline{\psfig{figure=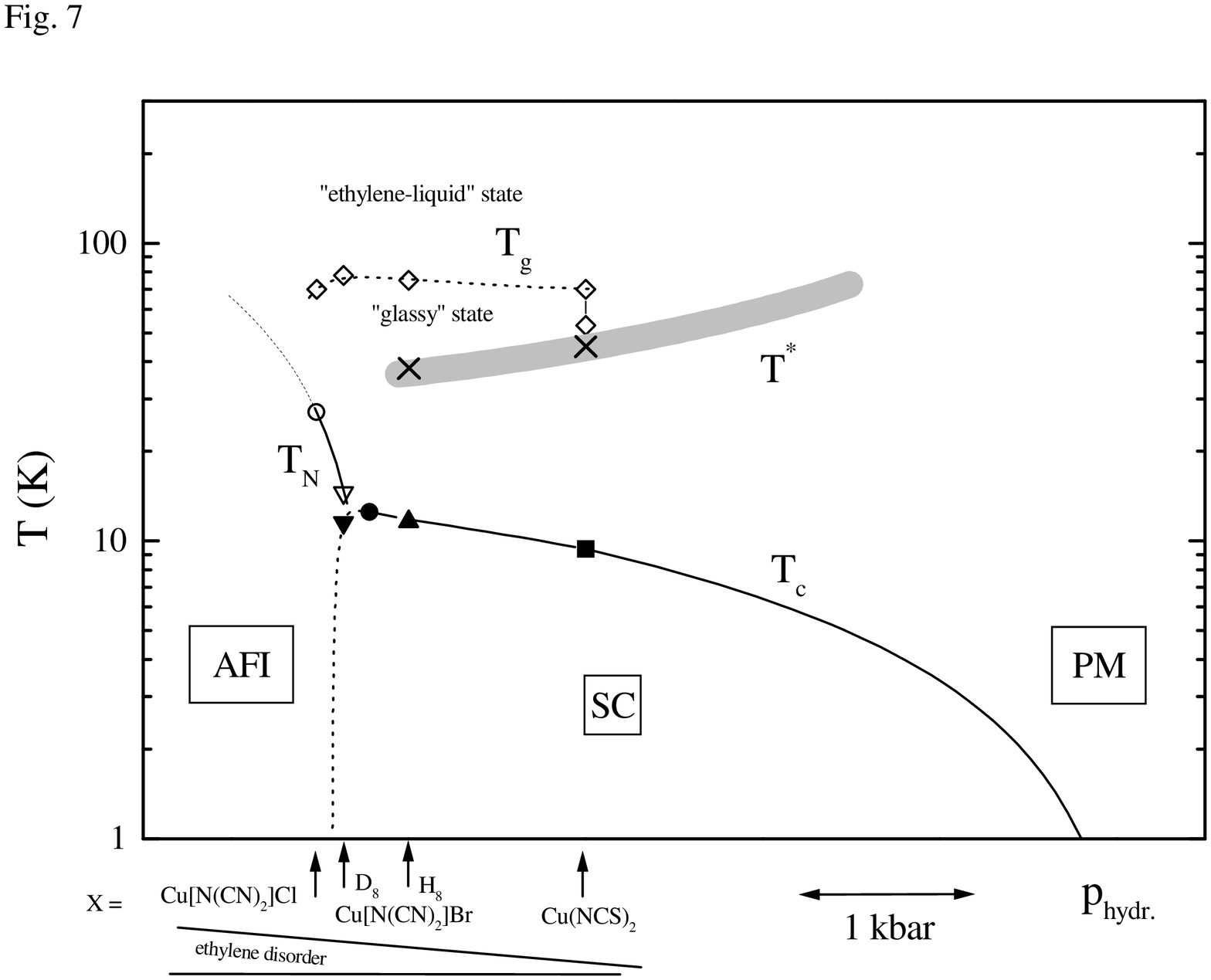,width=\textwidth}}
\vspace{0.5cm} \caption{Temperature/hydrostatic-pressure phase
diagram for the $\protect\kappa$-(BEDT-TTF)$_2$X title compounds.
Arrows indicate the location of different compounds at ambient
pressure. Solid lines represent the hydrostatic pressure
dependencies of $T_N$ and $T_c$ taken from literature. Circles
denote $\protect\kappa$-(ET)$_2$Cu[N(CN)$_2$]Cl, down and up
triangles denote deuterated and hydrogenated
$\protect\kappa$-(ET)$_2$Cu[N(CN)$_2$]Br, respectively, and
squares stand for $\protect\kappa$-(ET)$_2$Cu(NCS)$_2$. The
superconducting and antiferromagnetic transitions are represented
by closed and open symbols, respectively. Diamonds stand for the
glass-like transitions and crosses for the maxima of the anomalous
expansivity contributions $\protect\delta \protect\beta$ at
intermediate temperatures $30\,{\rm K}-50$\,K. In the shaded area
anomalies in magnetic, transport and elastic properties have been
observed (details are explained in the text).} \label{phase}
\end{figure}

Fig.\ \ref{phase} summarizes our results on the
various $\kappa $-(ET)$_{2}$X compounds discussed in the present
paper. To this end we use as the abscissa hydrostatic pressure
following the suggestion of \cite{Kanoda97}. The positions of the
various salts at ambient pressure is indicated by the arrows. The
solid lines representing the phase boundaries from the
paramagnetic (PM) to the superconducting (SC) and insulating
antiferromagnetic (AFI) states refer to the results of
hydrostatic-pressure studies of $T_{c}$ and $T_{N}$
\cite{Schirber88,Schirber90,Schirber91}. It is important to keep
in mind, however, that the pressure effects are highly
anisotropic. In an attempt to find out the relevant directional
dependent material-to-properties correlations, no simple
systematics was found for the {\em intra}plane pressure
coefficients of $T_{c}$. On the other hand, a feature common to
all systems studied among the $\kappa $-(ET)$_{2}$X salts is the
large negative effect for uniaxial stress perpendicular to the
planes on both $T_{c}$ as well as $T_{N}$. In both cases the
pronounced {\em interlayer} effect dominates the $T_{c}$ and
$T_{N}$ shifts found under hydrostatic-pressure conditions. Our
results provide clear thermodynamic evidence for a second-order
phase transition at $T_{N}= 27.8$\,K in $\kappa $-Cl.

At elevated temperatures, a glass transition at $T_g$ has been
identified that defines the boundary between an ethylene-liquid at
$T>T_g$ and a glassy state at $T<T_g$ (dotted line): at
temperatures above $T_g$, the motional degrees of freedom of the
ethylene endgroups can be excited and thus contribute to the
specific heat and thermal expansion whereas below $T_g$ a certain
disorder is frozen in. The glass-like transition which is
structural in nature has been shown to imply time dependencies in
electronic properties. We discussed possible implications on the
ground-state properties of $\kappa$-Br in terms of frozen-in
lattice disorder depending on the cooling rate employed at
$T_{g}$. Using similar cooling conditions, the degree of frozen-in
disorder seems to be largest for $\kappa$-Cl and decreases from
$\kappa$-Br to $\kappa$-Cu(NCS)$_{2}$. Considering the large
pressure/volume dependence of $S_{{\rm ethy}}$ for the $\kappa$-Br
salt we proposed a pressure experiment which could clarify the
relative role of lattice disorder for the ground-state properties
of hydrogenated and deuterated $\kappa$-Br.

At intermediate temperatures, we observe an anomalous expansivity
contribution, $\delta \beta (T)$, at $T^{\ast}$ for the
superconducting salts, symbolized in Fig.\ \ref{phase} by crosses.
These anomalies coincide with various features observed in
magnetic, transport and elastic properties (shaded area). Among
them are the peak in $(T_{1}T)^{-1}$ and the decrease of
$\chi_{{\rm Spin}}$ that point to strong antiferromagnetic
fluctuations and the decrease of the density of states at the
Fermi level.\\ We proposed that, instead of a pseudogap on the
whole Fermi surface as has been frequently discussed in the
literature, a real gap associated with a SDW on minor parts forms
below $T^{\ast}$. This scenario implies that the SDW and
superconductivity involve disjunct parts on the Fermi surface,
i.e.\ that the antiferromagnetic fluctuations compete with rather
than contribute to the superconducting instability. In a
generalized picture including also the AFI phase, the nature of
the magnetic state in $\kappa$-Cl might be that of an
antiferromagnetic order among nearest-neighbor spins, i.e.\ a
commensurate spin-density-wave.

Although electron correlations are important for the quasi-2D
organic superconductors, we have shown that a phase diagram solely
based on electronic degrees of freedom is not appropriate. This is
supported by recent theoretical investigations of the 2D Hubbard
model and its application to the $\kappa $-Br and $\kappa $-Cl
salts \cite{Painelli00}. It has been found that one {\em single}
parameter like the Coulomb repulsion $U_{{\rm eff}}$, the
effective transfer integrals $t_{{\rm eff}}$ or the orthorhombic
distortion $c/a$ cannot govern the physics of the AFI/SC
interface. It is a combination of electronic correlations,
electron-phonon coupling and interlayer effects as well as the
influence of lattice disorder which have to be considered in order
to understand more clearly the interesting physics of the $\kappa
$-(ET)$_{2}$X title compounds.

\subsection*{Acknowledgment}
We acknowledge fruitful discussions with F.\ Kromer, T.\ Cichorek,
B.\ Wolf, S.\ Zherlitsyn, A.\ Goltsev, C.\ Meingast, P.\ Nagel and
K.\ Maki. Work at Argonne National Laboratory is sponsored
supported by the U.S.\ Department of Energy, Office of Basic
Energy Sciences, Division of Materials Sciences, Divisions of
Materials Science, under contract W-31-109-ENG-38.

\onecolumn
\pagebreak

\begin{references}
\bibitem{Leung85} P.\ C.\ W.\ Leung, T.\ J.\ Emge, M.\ A.\ Beno, H.\ H.\
Wang, J.\ M.\ Williams, V.\ Petricek, and P.\ Coppens, J. Am. Chem. Soc.
{\bf 107}, 6184 (1985).

\bibitem{Geiser91} U.\ Geiser, A.\ J.\ Schultz, H.\ H.\ Wang, D.\ M.\
Watkins, D.\ L.\ Stupka, J.\ M.\ Williams, J.\ E.\ Schirber, D.\ L.\
Overmyer, D.\ Jung, J.\ J.\ Novoa, and M.-H.\ Whangbo, Physica C {\bf 174},
475 (1991).

\bibitem{Ishiguro97} T.\ Ishiguro, K.\ Yamaji, and G.\ Saito, Organic
Superconductors, Second edition, Springer, Berlin (1997).

\bibitem{Lang96} M.\ Lang, Superconductivity Review {\bf 2}, 1 (1996).

\bibitem{McKenzie97} R.\ H.\ McKenzie, Science {\bf 278}, 820 (1997).

\bibitem{Wang91} H.\ H.\ Wang, K.\ D.\ Karlson, U.\ Geiser, A.\ M.\ Kini,
A.\ J.\ Schultz, J.\ M.\ Williams, L.\ K.\ Montgomery, W.\ K.\ Kwok, U.\
Welp, K.\ G.\ Vandervoort, S.\ J.\ Boryshuk, A.\ V.\ Strieby Crouch, J.\ M.\
Kommers, D.\ M.\ Watkins, J.\ E.\ Schirber, D.\ L.\ Overmyer, D.\ Jung, J.\
J.\ Novoa, and M.\ H.\ Whangbo, Synth. Met. {\bf 41-43}, 1983 (1991).

\bibitem{Kanoda97} K.\ Kanoda, Hyperfine Int. {\bf 104}, 235 (1997); K.\
Kanoda, Physica C {\bf 282-287}, 299 (1997).

\bibitem{Kino96} H.\ Kino and H.\ Fukuyama, J. Phys. Soc. Japan {\bf 65},
2158 (1996).

%\bibitem{Demiralp97} E.\ Demiralp and W.\ A.\ Goddard III, Phys.
%Rev. B {\bf 56}, 11907 (1997).

\bibitem{Nakazawa00} Y.\ Nakazawa, H.\ Taniguchi, A.\ Kawamoto, and K.\
Kanoda, Phys. Rev. B {\bf 61}, R16295 (2000).

\bibitem{Kawamoto95} A.\ Kawamoto, K.\ Miyagawa, Y.\ Nakazawa, and K.\
Kanoda, Phys. Rev. Lett. {\bf 74}, 3455 (1995).

\bibitem{Mayaffre94} H.\ Mayaffre, P.\ Wzietek, C.\ Lenoir, D.\, J$\acute{%
{\rm e}}$rome, and P.\ Batail, Europhys. Lett. {\bf 28}, 205
(1994).

\bibitem{Kawamoto95a} A.\ Kawamoto, K.\ Miyagawa, Y.\ Nakazawa, and K.\
Kanoda, Phys. Rev. B {\bf 52}, 15522 (1995).

\bibitem{Sushko91} Y.\ V.\ Sushko, V.\ A.\ Bondarenko, R.\ A.\ Petrosov, N.\
D.\ Kushch, and E.\ B.\ Yagubskii, J. Phys. I France {\bf 1} 1375 (1991).

\bibitem{Gärtner88} S.\ G\"artner, E.\ Gogu, I.\ Heinen, H.\ J.\ Keller, T.\
Klutz, and D.\ Schweitzer, Solid State Commun. {\bf 65}, 1531 (1988).

\bibitem{Murata90} K.\ Murata, M.\ Ishibashi, Y.\ Honda, N.\ A.\
Fortune, M.\ Tokumoto, N.\ Kinoshita, and H.\ Anzai, Solid State
Commun. {\bf 76}, 377 (1990).

\bibitem{Kataev92} V.\ Kataev, G.\ Winkel, D.\ Khomskii, D.\ Wohlleben, W.\
Crump, K.\ F.\ Tebbe, and J.\ Hahn, Solid State Commun. {\bf 83}, 435 (1992).

\bibitem{Wzietek96} P.\ Wzietek, H.\ Mayaffre, D.\ J$\acute{{\rm e}}$rome,
and S.\ Brazovskii, J. Phys. I France {\bf 6}, 2011 (1996).

\bibitem{Frikach00} K.\ Frikach, M.\ Poirier, M.\ Castonguay, and K.\ D.\
Truong, Phys. Rev. B {\bf 61}, R6491 (2000).

\bibitem{Kawamoto97} A.\ Kawamoto, K.\ Miyagawa, and K.\ Kanoda, Phys. Rev.
B {\bf 55}, 14140 (1997).

\bibitem{Su98a} X.\ Su, F.\ Zuo, J.\ A.\ Schlueter, M.\ E.\ Kelly, and J.\
M.\ Williams, Phys. Rev. B {\bf 57}, R14056 (1998).

\bibitem{Su98b} X.\ Su, F.\ Zuo, J.\ A.\ Schlueter, M.\ E.\ Kelly, and J.\
M.\ Williams, Solid State Commun. {\bf 107}, 731 (1998).

\bibitem{Su98c} X.\ Su, F.\ Zuo, J.\ A.\ Schlueter, A.\ M.\ Kini, and J.\
M.\ Williams, Phys. Rev. B {\bf 58}, R2944 (1998).

\bibitem{Ito00} H.\ Ito, T.\ Ishiguro, T.\ Kondo, and G.\ Saito, Phys. Rev.
B {\bf 61}, 3243 (2000).

\bibitem{Kini96} A.\ M.\ Kini, K.\ D.\ Carlson, H.\ H.\ Wang, J.\ A.\
Schlueter, J.\ D.\ Dudek, S.\ A.\ Sirchio, U.\ Geiser, K.\ R.\ Lykke, and
J.\ M.\ Williams, Physica C {\bf 264}, 81 (1996).

\bibitem{Pintschovius97} L.\ Pintschovius, H.\ Rietschel, T.\ Sasaki, H.\
Mori, S.\ Tanaka, N.\ Toyota, M.\ Lang, and F.\ Steglich, Europhys. Lett.
{\bf 37}, 627 (1997).

\bibitem{Elsinger00} H.\ Elsinger, J.\ Wosnitza, S.\ Wanka, J.\ Hagel, D.\
Schweitzer, and W.\ Strunz, Phys. Rev. Lett. {\bf 84}, 6098 (2000).

\bibitem{DeSoto95} S.\ M.\ De\,Soto, C.\ P.\ Slichter, A.\ M.\ Kini, H.\ H.\
Wang, U.\ Geiser, and J.\ M.\ Williams, Phys. Rev. B {\bf 52}, 10364 (1995).

\bibitem{Mayaffre95} H.\ Mayaffre, P.\ Wzietek, D.\ J$\acute{{\rm e}}$rome,
C.\ Lenoir, and P.\ Batail, Phys. Rev. Lett. {\bf 75}, 4122 (1995).

\bibitem{Kanoda96} K.\ Kanoda, K.\ Miyagawa, A.\ Kawamoto, and Y.\ Nakazawa,
Phys. Rev. B {\bf 54}, 76 (1996).

\bibitem{Wosnitza99} J.\ Wosnitza, J. Low Temp. Phys. {\bf 117}, 1701 (1999).

\bibitem{Mueller00} J.\ M\"uller, M.\ Lang, F.\ Steglich, J.\ A.\ Schlueter,
A.\ M.\ Kini, U.\ Geiser, J.\ Mohtasham, R.\ W.\ Winter, G.\ L.\ Gard, T.\
Sasaki, and N.\ Toyota, Phys. Rev. B {\bf 61}, 11739 (2000).

\bibitem{Mueller01} J.\ M\"uller, M.\ Lang, J.\ A.\ Schlueter, U.\ Geiser,
and D.\ Schweitzer, Synth. Met. {\bf 120}, 855 (2001).

\bibitem{Pott83} R.\ Pott and R.\ Schefzyk, J. Phys. E {\bf 16}, 445 (1983).

\bibitem{Wang90} H.\ H.\ Wang, A.\ M.\ Kini, L.\ K.\ Montgomery, U.\ Geiser,
K.\ D.\ Carlson, J.\ M.\ Williams, J.\ E.\ Thompson, D.\ M.\ Watkins, W.\
K.\ Kwok, U.\ Welp, and K.\ G.\ Vandervoort, Chem. Mat. {\bf 2}, 284 (1990).

\bibitem{Kini96b} A.\ M.\ Kini, H.\ H.\ Wang, J.\ A.\ Schlueter,
J.\ D.\ Dudek, U.\ Geiser, K.\ D.\ Carlson, J.\ M.\ Williams, M.\
E.\ Kelly, E.\ Stevenson, A.\ S.\ Komosa, S.\ A.\ Sirchio, Mol.
Cryst. Liq. Cryst. {\bf 284}, 419 (1996).

\bibitem{Kund95} M.\ Kund, J.\ Lehrke, W.\ Biberacher, A.\ Lerf, and K.\
Andres, Synth. Met. {\bf 70}, 949 (1995).

\bibitem{Miyagawa95} K.\ Miyagawa, A.\ Kawamoto, Y.\ Nakazawa, and K.\
Kanoda, Phys. Rev. Lett. {\bf 75}, 1174 (1995).

%\bibitem{Sushko93} Y.\ V.\ Sushko, and K.\ Andres, Phys. Rev. B {\bf
%47}, 330 (1993).

\bibitem{Gugenberger92} F.\ Gugenberger, R.\ Heid, C.\ Meingast, P.\
Adelmann, M.\ Braun, H.\ W\"uhl, M.\ Haluska, and H.\ Kuzmany, Phys. Rev.
Lett. {\bf 69}, 3774 (1992).

\bibitem{Cahn91} for a review, see, e.g.: R.W.\ Cahn (editor), Glasses and
Amorphous Materials, Materials Science and Technology {\bf 9}, Verlag
Chemie, Weinheim, 137 (1991); W.\ Kauzmann, The nature of the glassy state
and the behavior of liquids at low temperatures, Chem. Rev. {\bf 43}, 219
(1948).

\bibitem{Angell95} C.A.\ Angell, Science {\bf 267}, 1924 (1995).

\bibitem{Saito99} K.\ Saito, H.\ Akutsu, and M.\ Sorai, Solid State Commun.
{\bf 111}, 471 (1999).

\bibitem{Akutsu00} H.\ Akutsu, K.\ Saito, and M.\ Sorai, Phys. Rev. B {\bf 61%
}, 4346 (2000).

\bibitem{Kund93} M.\ Kund, H.\ M\"uller, W.\ Biberacher, and K.\ Andres,
Physica C {\bf 191}, 274 (1993).

\bibitem{Kund94} M.\ Kund, K.\ Andres, H.\ M\"uller, and G.\ Saito, Physica
B {\bf 203}, 129 (1994).

\bibitem{deBolt76} M.\ A.\ deBolt, A.\ J.\ Easteal, P.\ B.\ Macedo, and C.\
T.\ Moynihan, J. Am. Ceram. Soc. {\bf 59}, 16 (1976).%
%; C.\ A.\ Angell, J. Non-Cryst. Solids {\bf 131-133}, 13
%(1991).

\bibitem{Nagel00} P.\ Nagel, V.\ Pasler, C.\ Meingast, A.\ I.\ Rykov, and
S.\ Tajima, Phys. Rev. Lett. {\bf 85}, 2376 (2000).

\bibitem{comment1a} A two-state model consists of a ground state and an
excited state which are separated by the {\em energy difference}
$E_{{\rm S}}$. Additionally an {\em energy barrier} $E_{{\rm a}}$
between both states is assumed.

\bibitem{Rahal97} M.\ Rahal, D.\ Chasseau, J.\ Gaultier, L.\ Ducasse, M.\
Kurmoo, and P.\ Day, Acta Cryst. B {\bf 53}, 159 (1997). %; D.\
%Chasseau, J.\ Gaultier, M.\ Rahal, L.\ Ducasse, M.\ Kurmoo, and
%P.\ Day, Synth. Met. {\bf 41-43}, 2039 (1991);

\bibitem{Tanatar99} M.\ A.\ Tanatar, T.\ Ishiguro, T.\ Kondo, and G.\ Saito,
Phys. Rev. B {\bf 59}, 3841 (1999).

\bibitem{Kini90} A.\ M.\ Kini, U.\ Geiser, H.\ H.\ Wang, K.\ D.\ Carlson,
J.\ M.\ Williams, W.\ K.\ Kwok, K.\ G.\ Vandervoort, J.\ E.\ Thompson, D.\
L.\ Stupka, D.\ Jung, and M.-H.\ Whangbo, Inorg. Chem. {\bf 29}, 2555 (1990).

\bibitem{Geiser91a} U.\ Geiser, A.\ M.\ Kini, H.\ H.\ Wang, M.\ A.\ Beno,
and J.\ M.\ Williams, Acta Cryst. C {\bf 47}, 190 (1991).

\bibitem{comment2} These values should not be compared with the jump heights
$\Delta \alpha _{i}$ reported by Kund {\em et al.}
\cite{Kund93,Kund94} where the authors treated the anomalies as
second-order phase transitions, thereby using an "equal-areas"
construction to replace the heating curves by an idealized sharp
jump. However, as we have shown, the shape of the heating curves
is not unique but depends on the thermal history. The jump of the
volume expansivity, $\Delta \beta $, shown in Fig.\ 4 of Ref.
\cite{Kund93} can clearly be attributed to the use of this
improper procedure. However, looking closely at their data apart
from the under- and overshoot behavior in $\alpha _{i}(T)$ one
finds a reasonable agreement with the values reported here. %Furthermore,
%comparing our estimated values of $\Delta \beta(T_g)$ for the
%$\kappa$-Br and $\kappa$-Cl salts to the previously reported
%thermal expansion measurements \cite{Kund93,Kund94}, we find the
%same value for the ratio $\Delta \beta$($\kappa$-Br) / $\Delta
%\beta$($\kappa$-Cl) suggesting that the way of the analysis of our
%data is correct.

%\bibitem{Barron80} T.\ H.\ Barron, J.\ G.\ Collins, and G.\ K.\ White,
%Advances in Physics {\bf 29}, 609 (1980).

%\bibitem{Sato00} A.\ Sato, H.\,Akutsu, K.\ Saito, and M.\ Sorai,
%Proceedings ICSM 2000, Bad Gastein, in press.

%\bibitem{comment3} As already pointed out in \cite{comment2}, the
%estimated values of $\Delta \beta(T_g)$ for the $\kappa$-Br and
%$\kappa$-Cl salts differ from the values reported by Kund {\em et
%al.} because the authors didn't consider the peculiarities of a
%glass-like transition. Nevertheless, comparing our data for both
%salts to the previously reported thermal expansion measurements
%\cite{Kund93,Kund94}, we find the same value for the ratio $\Delta
%\beta$($\kappa$-Br) / $\Delta \beta$($\kappa$-Cl) suggesting that
%the way of the analysis of our data is correct.

\bibitem{Whangbo90} M.\ H.\ Whangbo, J.\ J.\ Novoa, D.\ Jung, J.\ M.\
Williams, A.\ M.\ Kini, H.\ H.\ Wang, U.\ Geiser, M.\ A.\ Beno, and K.\ D.\
Carlson, Organic Superconductivity, ed. by V.\ Z.\ Kresin and W.\ A.\
Little, Plenum Press, New York, 243 (1990).

\bibitem{Watanabe91} Y.\ Watanabe, H.\ Sato, T.\ Sasaki, and N.\ Toyota, J.
Phys. Soc. Japan {\bf 60}, 3608 (1991).

\bibitem{Aburto98} A.\ Aburto, L.\ Fruchter, and C.\ Pasquier, Physica C
{\bf 303}, 185 (1998).

%\bibitem{Taniguchi99} H.\ Taniguchi, A.\ Kawamoto, and K.\ Kanoda,
%Phys. Rev. B {\bf 59}, 8424 (1999).

\bibitem{Nakazawa97} Y.\ Nakazawa and K.\ Kanoda, Phys. Rev. B {\bf 55},
R8670 (1997).

\bibitem{Kohno99} H.\ Kohno, H.\ Fukuyama, and M.\ Sigrist, J. Phys. Soc.
Japan {\bf 68}, 1500 (1999).

%\bibitem{Lin98} Y.\ Lin, J.\ E.\ Eldridge, H.\ H.\ Wang, A.\ M.\
%Kini, M.\ E.\ Kelly, J.\ M.\ Williams, and J.\ A.\ Schlueter,
%Phys. Rev. B {\bf 58}, R599 (1998).

\bibitem{Welp92} U.\ Welp, S.\ Fleshler, W.\ K.\ Kwok, G.\ W.\ Crabtree, K.\
D.\ Carlson, H.\ H.\ Wang, U.\ Geiser, J.\ M.\ Williams, and V.\ M.\
Hitsman, Phys. Rev. Lett. {\bf 69}, 840 (1992).

%\bibitem{Schultz93} A.\ J.\ Schultz, U.\ Geiser, H.\ H.\ Wang, J.\
%M.\ Williams, L.\ W.\ Finger, and R.\ M.\ Hazen, Physica C {\bf
%208}, 277 (1993).

\bibitem{Posselt94} H.\ Posselt, H.\ M\"uller, K.\ Andres, and G.\ Saito,
Phys. Rev. B {\bf 49}, 15849 (1994).

\bibitem{Tanatar97} M.\ A.\ Tanatar, T.\ Ishiguro, H.\ Ito, M.\ Kubota, and
G.\ Saito, Phys. Rev. B {\bf 55}, 12529 (1997).

\bibitem{comment4} As mentioned in section \ref{glas1}, due to the shape of
our $\kappa$-(ET)$_2$Cu[N(CN)$_2$]Cl single crystal only the
diagonal in-plane axis $\alpha_d = \frac{1}{2} \times (\alpha_a +
\alpha_c)$ could be measured. Since thermal expansion results on
the same system performed by Kund {\em et al.} \cite{Kund94}
reveal no discontinuity neither for $\alpha_a$ nor for $\alpha_c$,
an accidental cancellation of the two intralayer effects, i.e.\
$\Delta \alpha_a = - \Delta \alpha_c$ can be ruled out.

\bibitem{Nakazawa96} Y.\ Nakazawa, and K.\ Kanoda, Phys. Rev. B {\bf 53},
R8875 (1996).

\bibitem{Schirber91} J.\ E.\ Schirber, D.\ L.\ Overmyer, K.\ D.\ Carlson,
J.\ M.\ Williams, A.\ M.\ Kini, H.\ H.\ Wang, H.\ A.\ Charlier, B.\ J.\
Love, D.\ M.\ Watkins, and G.\ A.\ Yaconi, Phys. Rev. B {\bf 44}, 4666
(1991).

\bibitem{Lefebvre00} S.\ Lefebvre, P.\ Wzietek, S.\ Brown, C.\ Bourbonnais,
D.\ J$\acute{{\rm e}}$rome, C.\ M$\acute{{\rm e}}$zi$\grave{{\rm e}}$re, M.\
Fourmigu$\acute{{\rm e}}$, and P.\ Batail, Phys. Rev. Lett. {\bf 85}, 5420
(2000).

\bibitem{Sun91} K.\ Sun, J.\ H.\ Cho, F.\ C.\ Chou, W.\ C.\ Lee, L.\ L.\
Miller, D.\ C.\ Johnston, Y.\ Hidaka, and T.\ Murakami, Phys. Rev.
B {\bf 91}, 239 (1991).

\bibitem{Siurakshina00} L.\ Siurakshina, D.\ Ihle, and R.\ Hayn,
Phys. Rev. B {\bf 61}, 14601 (2000).

\bibitem{McKenzie98} R.\ H.\ McKenzie, Comments Cond. Mat. Phys.
{\bf 18}, 309 (1998).

\bibitem{Maesato01} M.\ Maesato, Y.\ Kaga, R.\ Kondo, H.\ Hirai, and S.\
Kagoshima, Synth. Met. {\bf 120}, 941 (2001).

\bibitem{Campos95}C.\ E.\ Campos, J.\ S.\ Brooks, P.\
J.\ M.\ van Bentum, J.\ A.\ A.\ J.\ Perenboom, S.\ J.\ Klepper,
P.\ S.\ Sandhu, S.\ Valfells, Y.\ Tanaka, T.\ Kinoshita, N.\
Kinoshita, M.\ Tokumoto, H.\ Anzai, {\em Phys. Rev. B} {\bf 52}
R7014 (1995).

\bibitem{Kund94a} M.\ Kund, H.\ Veith, H.\ M\"uller, K.\ Andres, and G.\
Saito, Physica C {\bf 221}, 119 (1994).

\bibitem{Elsinger00a} We used the value for the specific heat jump reported
in \cite{Elsinger00}. These authors applied a BCS fit with strong
coupling to their data giving $\Delta C / \gamma T_c = 3.3$ and
$\Delta C \approx 0.95 $\,J/(mol\ K). A value of $\Delta C \geq
0.8$\,J/(mol\ K) can be directly read off the data; nevertheless,
an uncertainty of $10\,\%\ ...\ 20\,\%$, must be considered [J.\
Wosnitza, private communication]. The very good agreement of the
calculated hydrostatic-pressure coefficient of $T_c$ with the
values found in pressure experiments suggests that using the above
value
of $\Delta C$ is reasonable rather than using $\Delta C / T = (45 \pm 10)$%
\,mJ/(mol\ K$^2$) at $T_c \sim 11.5\,{\rm K}$ [B.\ Andraka {\em et al.},
Solid State Commun. {\bf 79}, 57 (1991)] which would give a much higher
value for $\partial T_c / \partial p_{hydr}$.

\bibitem{comment5} Note that for the determination of $\partial T_c /
\partial p_{hydr}$ of the $\kappa$-Br salt the jump heights $\Delta \alpha_a$
and $\Delta \alpha_c$ of crystal \#\,3 and $\Delta \alpha_b$ of crystal
\#\,2 were used.

\bibitem{Schirber90} J.\ E.\ Schirber, D.\ L.\ Overmyer, J.\ M.\ Williams,
A.\ M.\ Kini, and H.\ H.\ Wang, Physica C {\bf 170}, 231 (1990).

%\bibitem{Scripov92} A.\ V.\ Scripov, and A.\ P.\ Stepanov, Physica
%C {\bf 197}, 89 (1992).

%\bibitem{Doi91} T.\ Doi, K.\ Oshima, H.\ Yamazaki, H.\ Murayama, H.\ Maeda,
%A.\ Koizumi, H.\ Kimura, M.\ Fujita, Y.\ Yunoki, H.\ Mori, S.\ Tanaka, H.\
%Yamochi, and G.\ Saito, J. Phys. Soc. Japan {\bf 60}, 1441 (1991).

%\bibitem{Girlando00} A.\ Girlando, M.\ Masino, G.\ Visentini, R.\ G.\ D.\
%Valle, A.\ Brillante, and E.\ Venuti, Phys. Rev. B {\bf 62}, 14476 (2000).

\bibitem{Simizu00} T.\ Simizu, N.\ Yoshimoto, M.\ Nakamura, and M.\
Yoshizawa, Physica B {\bf 281 \& 282}. 896 (2000).

\bibitem{unpublished} M.\ K\"oppen, M.\ Lang, and F.\
Steglich, unpublished results.

\bibitem{Sasaki01} T.\ Sasaki {\em et al.}, in preparation.

\bibitem{Kornelsen92} K.\ Kornelsen, J.\ E.\ Eldridge, H.\ H.\
Wang, H.\ A.\ Charlier, and J.\ M.\ Williams, Solid State Commun.
{\bf 81}, 343 (1992).

\bibitem{Schirber88} J.\ E.\ Schirber, E.\ L.\ Venturini, A.\ M.\ Kini, H.\
H.\ Wang, J.\ R.\ Witworth and J.\ M.\ Williams, {\em Physica C}
{\bf 152}, 157 (1988).

\bibitem{Painelli00} A.\ Painelli, A.\ Girlando, and A.\ Fortunelli,
cond-mat/0011154, (2000).

\end{references}
\end{document}